\documentclass[iop]{emulateapj}
\usepackage{apjfonts}
\journalinfo{The Astrophysical Journal Supplement Series, 2014, in press}

\shorttitle{Proton Cyclotron Heating in the Solar Wind}
\shortauthors{S.~R.~Cranmer}

\begin{document}

\title{Ensemble Simulations of Proton Heating in the Solar Wind via
Turbulence and Ion Cyclotron Resonance}

\author{Steven R. Cranmer}

\affil{Harvard-Smithsonian Center for Astrophysics,
60 Garden Street, Cambridge, MA 02138, USA} 

\begin{abstract}
Protons in the solar corona and heliosphere exhibit anisotropic
velocity distributions, violation of magnetic moment conservation,
and a general lack of thermal equilibrium with the other particle
species.
There is no agreement about the identity of the physical processes
that energize non-Maxwellian protons in the solar wind, but a
traditional favorite has been the dissipation of ion cyclotron
resonant Alfv\'{e}n waves.
This paper presents kinetic models of how ion cyclotron waves heat
protons on their journey from the corona to interplanetary space.
It also derives a wide range of new solutions for the relevant
dispersion relations, marginal stability boundaries, and nonresonant
velocity-space diffusion rates.
A phenomenological model containing both cyclotron damping and
turbulent cascade is constructed to explain the suppression of
proton heating at low alpha--proton differential flow speeds.
These effects are implemented in a large-scale model of proton
thermal evolution from the corona to 1 AU.
A Monte Carlo ensemble of realistic wind speeds, densities,
magnetic field strengths, and heating rates produces a filled
region of parameter space (in a plane described by the parallel
plasma beta and the proton temperature anisotropy ratio) similar
to what is measured.
The high-beta edges of this filled region are governed by
plasma instabilities and strong heating rates.
The low-beta edges correspond to weaker proton heating and a
range of relative contributions from cyclotron resonance.
On balance, the models are consistent with other studies that
find only a small fraction of the turbulent power spectrum needs
to consist of ion cyclotron waves.
\end{abstract}


\section{Introduction}
\label{sec:intro}

The Sun's high-temperature corona expands into the heliosphere
as a supersonic, magnetized, and weakly collisional solar wind.
Despite many years of study, there is still no comprehensive
understanding of the physical processes that generate this
highly energized state.
Furthermore, it is unclear to what extent the solar wind detected
in interplanetary space preserves sufficient information from the
corona to help us learn how the plasma was heated initially.
It has been known for several decades that elemental abundances
and ionization fractions measured at 1~AU must be ``frozen in''
at low heights in the solar atmosphere \citep[e.g.,][]{Zu07}.
However, the temperatures and detailed velocity distribution functions
(VDFs) of ions and electrons appear to evolve gradually through
the heliosphere \citep{Me12} and in some cases they may be affected
by instabilities that become activated near the detecting spacecraft
\citep{Ga01}.

What are the processes that affect the thermodynamics of positive
ions as they accelerate away from the solar corona?
Because of infrequent Coulomb collisions above the coronal base,
particles that flow along magnetic field lines should want to
conserve their magnetic moments \citep{CGL}.
\citet{HS68} found that without any other source of heating,
magnetic moment conservation would produce extremely cold and
beamed (in the parallel sense; $T_{\parallel} \gg T_{\perp}$)
heliospheric protons, which is not seen \citep[e.g.,][]{Mr12}.
In fact, at 1~AU the majority of proton VDFs are close to isotropic,
which seems to require some residual or anomalous coupling via
collisions \citep{GD69,LM87}.
Heat conduction is an important carrier of thermal energy for
electrons, but not so much for protons \citep{SL95}.
The dominant sources of thermal energy for protons are
believed to be the irreversible decays of plasma structures;
i.e., dissipation of waves, shredding of turbulent eddies, and
multiple types of energy conversion in current sheets associated
with magnetic reconnection.

The goal of this paper is to determine the consequences of one
specific proposed idea for proton heating in the solar wind:
the dissipation of ion cyclotron resonant waves.
Although other sources of heat are likely to exist, we find it useful
to explore how much can be explained by restricting ourselves to
just a single main process.
This mechanism has been studied extensively
\citep[see reviews by][]{HI02,Ma06}, but often the microphysics of
wave-particle interactions are decoupled from the macrophysics of
VDF transport from the corona to 1~AU.
Thus, this paper aims to provide an in-depth study of how the
microphysics and macrophysics depend on one another.
Several well-known pieces of the puzzle will be assembled together
in new ways.

Section \ref{sec:summ} summarizes the wide range of suggested
physical explanations for ion heating in the solar wind and
lays out many of the open questions.
Section \ref{sec:disp} begins our focused look at the ion cyclotron
resonance mechanism by deriving several versions of the wave
dispersion relation.
Section \ref{sec:heat} discusses the net transfer of energy from
the waves to the anisotropic particle VDFs.
Section \ref{sec:alpha} presents a model of how drifting alpha
particles may suppress the cyclotron heating available to heliospheric
protons.
Section \ref{sec:radial} assembles the above results into a
large-scale model of radial energy transport from the corona
to 1~AU, and it shows how the observed distribution of states (in a
plane described by the parallel plasma beta and temperature anisotropy
ratios of protons) is explainable as a consequence of ion cyclotron
heating.
Section \ref{sec:conc} concludes with a discussion of some of the wider
implications of this work and gives suggestions for future improvements.

\section{A Walk Through the Ion Heating Maze}
\label{sec:summ}

Prior to delving into the primary physical process studied in
this paper, it is useful to list the alternatives.
Ideally, each of the proposed heating mechanisms should be tested
in a similar way as the ion cyclotron resonance idea is put through
its paces in Sections \ref{sec:disp}--\ref{sec:radial} below.
Here we are able to discuss only a fraction of the large number of
papers that presented ideas for the kinetic energization of ions
in the solar wind; for other reviews, see
\citet{Ho08}, \citet{Cr09}, and \citet{Of10}.
Also, since our goal is to study collisionless processes
that give rise to {\em preferential ion heating and acceleration,}
we neglect the much broader literature of strictly
magnetohydrodynamic (MHD) theories that do not focus on the kinetic
consequences of coronal heating \citep[see, e.g.,][]{Kl06,PD12}.

Particle and field instruments at heliocentric distances greater
than 0.3~AU have detected several marked departures from thermal
equilibrium for protons and other ions \citep{N82,Ma06,Kp08}.
In the fast solar wind, ions tend to be heated more strongly than
electrons, and protons often exhibit VDF anisotropies with
temperatures measured in the direction perpendicular to the
magnetic field often exceeding temperatures parallel to the field
(i.e., $T_{\perp} > T_{\parallel}$).
\citet{Ma83} found that proton magnetic moments are not conserved
between 0.3 and 1 AU in the fast wind; they increase steadily,
implying a gradual input of perpendicular kinetic energy.
Also, most heavy ion species flow faster than the protons by
about the local Alfv\'{e}n speed \citep{Hf98,Bg11}.
These measurements were augmented by spectroscopic observations of
similar extreme properties in low-density coronal holes
\citep{Ko97,Ko06,Wi11}.

Many of the models proposed to explain the proton and ion
measurements involve the damping of MHD waves.
However, there is little agreement about the most relevant wave
types, the dominant wave generation mechanisms, or the precise
means of dissipation.
Quite a few of the models also involve a broad array of multiple
steps of energy conversion between waves, turbulent eddies,
reconnection structures, and other nonlinear plasma features.
Nevertheless, it was noticed early on \citep{AF77,HT78}
that one specific mechanism---cyclotron resonance between left-hand
polarized Alfv\'{e}n waves and ion Larmor orbits---appears
to naturally produce many of the observed particle properties
\citep[see, e.g.,][]{Ma06,Ho08}.
Resonant ions ``surf'' along with a wave's oscillating electric
and magnetic fields, and they experience secular gains or losses
in energy depending on phase.
For a random distribution of wave phases, ions undergo a
diffusive random walk in velocity space \citep{KE66,Rw66}.

One major problem with the ion cyclotron model is that the resonant
wave frequencies $f$ in the corona are of order $10^2$ to $10^4$ Hz.
These frequencies are substantially higher than those inferred
for observed Alfv\'{e}n waves \citep[e.g., $f \lesssim 0.01$
Hz; see][]{Js09,Mc11}, and it is difficult find a process that bridges
that gap.
\citet{Ax92} and \citet{TM97} proposed that the low solar atmosphere
produces a continuous frequency spectrum of MHD waves that extends
up to $\sim$10$^{4}$ Hz.
As the waves propagate up into the corona, the high-frequency
end of the spectrum becomes eroded as the local Larmor frequency
decreases with increasing height.
However, \citet{Cr00,Cr01} argued that heavy ions with low
gyrofrequencies would likely be able to intercept nearly all of
the available wave energy prior to it becoming resonant with
protons and alpha particles.
Also, \citet{Ho00} found that the radio scintillation signature of
a basal spectrum of ion cyclotron waves would appear quite different
from what is already observed in the corona.
Thus, the idea that ion cyclotron waves are copiously generated
at the solar surface has fallen slightly out of favor,\footnote{%
There remains some uncertainty about the above criticisms of the
``basal generation'' idea.  Thus, definitive conclusions cannot
yet be made; see discussions in \citet{HI02} and \citet{Ma06}.}
and other explanations have been pursued.

A potentially more robust way to generate small-scale (i.e.,
high-frequency) plasma oscillations is to invoke a nonlinear
turbulent cascade.
Strong MHD turbulence is certainly present throughout the
heliosphere \citep[see reviews by][]{TM95,MV11,BCliving}.
Turbulent dissipation also appears able to provide the right
order of magnitude of heat to both the corona and solar wind
\citep[e.g.,][]{Dm02,CvB07,PC13,Li14}.
Many models invoke the idea that solar flux tubes are jostled
by photospheric granular motions, and this propels Alfv\'{e}n waves
into the corona that partially reflect back down to produce
counterpropagating wave packets.
Collisions of these wave packets are believed to drive an efficient
nonlinear cascade to small scales \citep{HN13}, but the ultimate
dissipation mechanisms are usually not identified in these models
\citep[see also][]{BP11,MS14}.

A distinguishing feature of MHD turbulence in the presence of a
strong background magnetic field is an anisotropic cascade in
wavenumber space.
Specifically, the breakup of eddies into smaller scales occurs
primarily in directions perpendicular to the field
\citep[e.g.,][]{St76,MT81,Sb83,GS95}.
The basic cascade process is {\em not} expected to produce
fluctuations with high parallel wavenumbers $k_{\parallel}$, and
thus it is not expected to directly excite much wave energy at
the ion cyclotron resonance.
Spacecraft have detected anisotropic turbulent power spectra in
interplanetary space that generally agree with these predictions
\citep{Hb08,Cn10,Sr10}, though there is also evidence that a
fraction of the fluctuation energy can be in the form of
high-frequency waves (see below).

If one treats the small-scale fluctuations in an MHD cascade as
linear waves, the dominant low-frequency modes at wavenumbers
$k_{\perp} \approx \rho_{p}^{-1}$, where $\rho_p$ is the proton
thermal gyroradius, are the kinetic Alfv\'{e}n wave (KAW) and
the magnetosonic whistler wave.
In collisionless plasmas, these modes tend to dissipate
via a combination of Landau and transit-time damping.
For conditions appropriate to the corona and heliosphere, the
KAW mode seems to be most prevalent \citep{Sc12,Pd13,Cn13,Rw13}.
Although linear KAW damping may one way of producing the fast
proton beams seen in interplanetary space \citep{VP13},
most of their energy goes into parallel electron heating
\citep{Lm99,CvB03,GB08} and not the perpendicular ion
heating that is observed.

For the last decade, there has been much work devoted to finding
the ways that low-frequency, high-$k_{\perp}$ turbulent fluctuations
can heat protons and heavy ions.
If KAW amplitudes become sufficiently high, the Larmor orbits of
protons and ions become stochastic.
This in turn enables low-energy particles to undergo random-walk
migration to higher energies, and this effectively produces
perpendicular heating \citep[e.g.,][]{JC01,VG04,Cs10,Ch11,Ch13,BC13}.
Alternately, KAW Landau damping may give rise to enough parallel
particle acceleration to produce beamed electron VDFs \citep{Hy14}
that themselves are unstable to the growth of Langmuir waves and
electrostatic ``phase space holes.''
The latter have been suggested as possible sources of perpendicular
scattering for protons and ions \citep{Mt03,CvB03}.
Still another idea is that turbulence may produce sufficiently strong
plasma inhomogeneities---i.e., velocity shears or cross-field density
gradients---that the system may become unstable to the rapid growth
of ion cyclotron waves \citep[see][]{Mk01,Mk06,VP08,Mi08,Ru12}.

It is possible that the smallest-scale fluctuations in MHD
turbulence are not accurately describable by a superposition of
linear waves.
Many simulations show that plasma turbulence eventually results
in the presence of intermittent vortices separated by
thin current sheets \citep[e.g.,][]{Ka13}.
Both test-particle models \citep{Dm04,Pa09,Lh09,Ly12}
and full Vlasov kinetic simulations \citep{Sv12,Sv14}
have shown that positive ions can interact resonantly with
turbulent current sheets and become heated perpendicularly.
A related idea is that when ions cross from the sub-Alfv\'{e}nic
reconnection inflow to the super-Alfv\'{e}nic outflow region,
they may undergo perpendicular acceleration via a rapid
pickup-like process \citep{Dr09,Ar14}.
There are also some ways that nonlinear Alfv\'{e}nic
fluctuations may drive different modes of ion VDF diffusion
than the ones predicted by classical linear or quasilinear wave
theory \citep[e.g.,][]{Mk09,DS13,Nk14}.

Despite the well-known tendency for MHD turbulence to be dominated
by a perpendicular cascade of low-frequency eddies
\citep[see also][]{Mt99,Hw08}, there is some
observational evidence for the existence of ion cyclotron waves.
There are time periods in which the solar wind exhibits
nearly monochromatic bursts of oscillation at or near the
local Larmor frequency \citep{Ts94,Jn09,Jn10}.
Additional empirical correlations between variance anisotropies,
the magnetic field geometry, and various helicity indices
indicate that a non-negligible fraction of the fluctuation
energy can be in the form of high-frequency Alfv\'{e}n waves
\citep{He11,Sm12}.
In the fast solar wind at $\sim$0.3 AU, non-Maxwellian shapes
of proton VDFs appear to be consistent with the velocity-space
diffusion that occurs in ion cyclotron wave dissipation
\citep{MT01b,Bu11}.

In addition to direct and indirect measurements of wave activity
at the ion cyclotron resonance, there are also quite a few
studies of the low-wavenumber inertial range
\citep[e.g.,][]{Bb96,Ds05,Mb08} that indicate the presence of
power-law spectra that extend to high values of {\em both}
$k_{\perp}$ and $k_{\parallel}$.
Even though the high-$k_{\parallel}$ spectra are not typically
seen to extend all the way up to ion cyclotron resonant wavenumbers,
they do hint at the existence of a kind of ``parallel cascade.''
Some models of MHD turbulence predict a weakened parallel
cascade that depends on higher-order wave-wave couplings than
the basic ones driving the perpendicular cascade
\citep{Ng96,Mv00,MG01}.
Other models propose that nonlinear couplings between Alfv\'{e}n
and fast-mode waves produce a high-$k_{\parallel}$ enhancement
in the Alfv\'{e}nic spectrum because of the nearly isotropic
cascade of the fast-mode waves \citep{Ch05,Ch08,CvB12}.

In this paper we will study the kinetic and thermodynamic
consequences of ion cyclotron resonant waves in the solar wind.
\citet{Ch10} and \citet{Is12} discussed several reasons why a
population of these waves may be dominated by nearly parallel
propagation (i.e., $k_{\parallel} \gg k_{\perp}$) if they are
present, and we restrict our analysis to this limiting wavenumber
condition as well.
We do not directly specify the origin of the ion cyclotron waves,
but merely note here that several of the mechanisms discussed
above---e.g., instabilities, multi-mode coupling, or a true
parallel cascade---may generate them gradually.
We do {\em not} require ion cyclotron waves to be the dominant
form of MHD fluctuation in the corona and solar wind, and
in fact the results presented below agree with earlier studies
\citep{IV11,CvB12} that found only a small fraction of the
total wave energy needs to be cyclotron resonant.
Thus, this work is largely consistent with other models in which
the turbulent cascade is dominated by low-frequency KAW-like
fluctuations.

\section{Parallel Alfv\'{e}n Wave Dispersion Relations}
\label{sec:disp}

Several key properties of the combined system of fluctuations and
background conditions depend on the dispersion relation of linear
MHD waves.
This section describes a number of different approaches that have
been used to describe the frequency of small-amplitude left-hand
polarized waves propagating parallel to a constant
magnetic field ${\bf B}_0$.
In general, the angular frequency $\omega$ is a complex function
of the real parallel wavenumber $k_{\parallel}$.
For reasons summarized above,
we assume a vanishingly small perpendicular wavenumber $k_{\perp}$,
which implies that the propagation angle
$\theta = \tan^{-1} (k_{\perp}/k_{\parallel}) \approx 0$.

For particle VDFs that remain gyrotropically symmetric around the
${\bf B}_0$ axis (and thus depend only on velocities $v_{\parallel}$
and $v_{\perp}$ parallel and perpendicular to the field, respectively),
the complex dielectric constant $\epsilon$ is given by
\begin{equation}
  \epsilon \, = \, 1 + \sum_{s}
  \frac{4\pi q_{s}^2}{m_{s} \omega^2} \int
  \frac{d^{3}{\bf v}}{\omega -k_{\parallel} v_{\parallel} -\Omega_s}
  \left( \frac{v_{\perp}}{2} \right)
  \hat{\Sigma} f_{s} ({\bf v})
  \label{eq:eorig}
\end{equation}
where the sum is taken over each species $s$ of particles having
masses $m_s$ and charges $q_s$ \citep[see, e.g.,][]{MT64,Stix}.
Each species' VDF is denoted by $f_{s} ({\bf v})$ and is normalized
to the total number density $n_s$ when integrated over all three
dimensions of velocity space.
In cylindrically symmetric (gyrotropic) coordinates,
$d^{3} {\bf v} = 2\pi v_{\perp} dv_{\perp} dv_{\parallel}$.
Each particle species also has its own Larmor gyrofrequency
$\Omega_{s} = q_{s} B_{0} / m_{s}c$, and the
pitch-angle derivative operator $\hat{\Sigma}$ is defined as
\begin{equation}
  \hat{\Sigma} \, = \, \left(
  \omega - k_{\parallel} v_{\parallel} \right)
  \frac{\partial}{\partial v_{\perp}} \, + \,
  k_{\parallel} v_{\perp} \frac{\partial}{\partial v_{\parallel}}
  \,\, .
\end{equation}
The solution to the dispersion relation is given by finding the
appropriate roots of
\begin{equation}
  \epsilon \, = \, k_{\parallel}^{2} c^{2} / \omega^2 \,\, .
  \label{eq:epseqn}
\end{equation}
In this paper the sum over $s$ is limited to three possible
species: electrons ($e$), protons ($p$), and alpha particles
($\alpha$).
The electrons are assumed to always follow the Maxwellian cold
plasma limit described below, and we assume the wave frequency
always obeys $| \omega / \Omega_{e} | \ll 1$.

\subsection{Bi-Maxwellian Velocity Distributions}
\label{sec:disp:bimax}

Many general properties of cyclotron resonant waves can be studied
in the simple limit of a bi-Maxwellian or two-temperature particle
distribution function.
For the proton VDF, we can assume
\begin{equation}
  f_{p} ( v_{\parallel}, v_{\perp} ) =
  \frac{n_p}{\pi^{3/2} w_{p \parallel} w_{p \perp}^2}
  \exp\! \left[ - \left( \frac{v_{\parallel} - u_{p \parallel}}
  {w_{p \parallel}} \right)^{2}
  - \left( \frac{v_{\perp}}{w_{p \perp}} \right)^{2} \right ]
  \label{eq:fbimax}
\end{equation}
where the proton number density is $n_p$ and the thermal spread
is described by temperatures parallel and perpendicular
to the field ($T_{p \parallel}$ and $T_{p \perp}$ respectively).
The corresponding thermal speeds are
\begin{equation}
  w_{p \parallel}^{2} = \frac{2k_{\rm B} T_{p \parallel}}{m_p}
  \,\, , \,\,\,\,\,
  w_{p \perp}^{2} = \frac{2k_{\rm B} T_{p \perp}}{m_p} \,\, ,
\end{equation}
with Boltzmann's constant given by $k_{\rm B}$ and the
anisotropy ratio defined as
${\cal R}_{p} = T_{p \perp} / T_{p \parallel}$.
The models below will be applied in the local rest frame of
the accelerating solar wind, which for a pure proton--electron plasma
implies that $u_{p \parallel} = 0$.

Even when using the simplified VDF of Equation (\ref{eq:fbimax}), there
are several levels of approximation that can be applied when
solving for the complex wave frequency $\omega = \omega_{r} + i \gamma$.
By convention, the linear fluctuations vary as $e^{-i\omega t}$, so
$\gamma > 0$ corresponds to unstable wave growth.
The most general way to solve Equation (\ref{eq:epseqn}) is to locate
all of its roots numerically for each desired wavenumber.
These solutions, often called solutions to the Vlasov--Maxwell
equations, have been presented extensively in the literature for
conditions relevant to the solar wind
\citep[e.g.,][]{Ga93,CvB03,CvB12,GB04,SY09,Yo10,Ma12}.

The limiting case of weak instability or decay
($| \gamma / \omega_{r} | \ll 1$) gives rise to some
interesting closed-form solutions of the dispersion relation.
Following, e.g., \citet{Dv83} and \citet{Bz93},
the dielectric constant can be written in this limit as
\begin{equation}
  \mbox{Re} (\epsilon) \, = \, 1 + \sum_{s}
  \left( \frac{\omega_{ps}}
  {\omega} \right)^{2} \, \left\{ \xi_{0} Z(\xi_{1}) +
  \left( {\cal R}_{s} - 1 \right) \left[ 1 +
  \xi_{1} Z(\xi_{1}) \right] \right\} \,\, ,
  \label{eq:reeps}
  \label{eq:epsR}
\end{equation}
\begin{equation}
  \mbox{Im} (\epsilon) \, = \, \sum_{s}
  \left( \frac{\omega_{ps}}{\omega} \right)^{2} \, \left\{
  \pi^{1/2} \exp ( -\xi_{1}^{2} ) \left[
  {\cal R}_{s} \xi_{1} +
  \frac{\Omega_s}{k_{\parallel} w_{\parallel}} \right] \right\} \,\, ,
  \label{eq:epsI}
\end{equation}
where the plasma frequencies are defined as
$\omega_{ps}^{2} = 4\pi q_{s}^{2} n_{s} / m_{s}$.
The dimensionless resonance parameter is given by
\begin{equation}
  \xi_{n} \, \equiv \,
  \frac{\omega_{r} - k_{\parallel} u_{s \parallel} - n \Omega_s}
  {k_{\parallel} w_{s \parallel}}
  \label{eq:xin}
\end{equation}
and the parallel drift speed $u_{s \parallel}$ is defined
relative to the local bulk (center of mass) speed of the plasma.
As above, we assume that $u_{p \parallel}=0$, but the value for
alpha particles may be nonzero.
The assumption of a Maxwellian VDF shape along any one direction is
encapsulated in the dimensionless plasma dispersion function, which
is defined as
\begin{equation}
  Z(\xi) \, \equiv \, \frac{1}{\sqrt{\pi}} \,
  {\cal P} \! \int_{-\infty}^{+\infty} \frac{dt \, e^{-t^2}}{t-\xi}
\end{equation}
where ${\cal P}$ denotes the Cauchy principal value
\citep[see][]{FC61}.

In the small-$\gamma$ limit, the real part of the dispersion relation
is simply 
$\mbox{Re}(\epsilon) = k_{\parallel}^{2} c^{2} / \omega_{r}^{2}$.
For convenience, we define dimensionless variables
\begin{equation}
  x \, \equiv \, \frac{k_{\parallel} V_{\rm A}}{\Omega_p}
  \,\, , \,\,\,\,\,\,
  y \, \equiv \, \frac{\omega_r}{\Omega_p}
  \label{eq:xydef}
\end{equation}
where $\omega_r$ is the real part of the frequency, and the
Alfv\'{e}n speed is given by
\begin{equation}
  V_{\rm A} \, = \, \frac{B_0}{\sqrt{4\pi\rho}}
  \,\, , \,\,\,\, \mbox{with} \,\,\,
  \rho \, = \, \sum_{s} m_{s} n_{s} \,\, .
\end{equation}
All of the solutions below make the nonrelativistic approximation
of $V_{\rm A} \ll c$, which is equivalent to ignoring the unity
term in Equation (\ref{eq:epsR}) and thus neglecting the displacement
current in Amp\`{e}re's law.

In the limit of small wavenumbers ($|x| \ll 1$), isotropic VDFs
(${\cal R} = 1$), and a pure proton--electron plasma, the dispersion
relation reduces to the ideal MHD expression for Alfv\'{e}n waves,
\begin{equation}
  y \, = \, x \,\, .
\end{equation}
If the small-wavenumber approximation is relaxed, the ``cold plasma''
dispersion relation for cyclotron waves can be derived by taking the
large-argument asymptotic limit for the dispersion function,
\begin{equation}
  Z(\xi) \, \approx \, -\frac{1}{\xi}  \,\, .
\end{equation}
Equation (\ref{eq:epsR}) is then solved to obtain
\begin{equation}
  x^{2} = \frac{y^2}{1-y} \,\, ,
  \label{eq:cdisp0}
\end{equation}
or
\begin{equation}
  y = \frac{x}{2} \left( \sqrt{x^{2} + 4} - x \right) 
  \label{eq:cdispP}
\end{equation}
\citep[see, e.g.,][]{Cp75,DH81,HI02}.
In this limiting case, the frequency does not depend on either
$T_{p \parallel}$ or $T_{p \perp}$.

The cold plasma dispersion relation has been generalized to include the
effect of additional ion species that drift relative to the protons.
Let us assume a relative alpha particle abundance
$h = n_{\alpha}/n_{p}$, and that the ions are flowing ahead of the
protons with a known value of
$\delta_{\alpha p} = (u_{\alpha \parallel} - u_{p \parallel})/V_{\rm A}$.
In the solar wind, $0 \leq \delta_{\alpha p} \leq 1$.
The ion drift necessitates a more precise accounting of the overall
charge neutrality and zero-current conditions in the plasma.
Following \citet{GE91}, \citet{HI02}, and others, the dispersion
relation becomes
\begin{equation}
  x^{2} = \frac{y^2}{1-y} +
  \frac{4h (y - x\delta_{\alpha p})^{2}}{1 - 2(y - x\delta_{\alpha p})}
  \,\, .
  \label{eq:cdispA}
\end{equation}
In general there are two separate branches of the dispersion
relation \citep[e.g.,][]{Is84a}.
Although there has been some work showing that both branches
may be excited in some situations \citep{GG94,Tu03},
for simplicity we assume that only the lowest
frequency branch is populated by wave power that cascades from
low-wavenumber fluctuations.

\begin{figure}
\epsscale{1.15}
\plotone{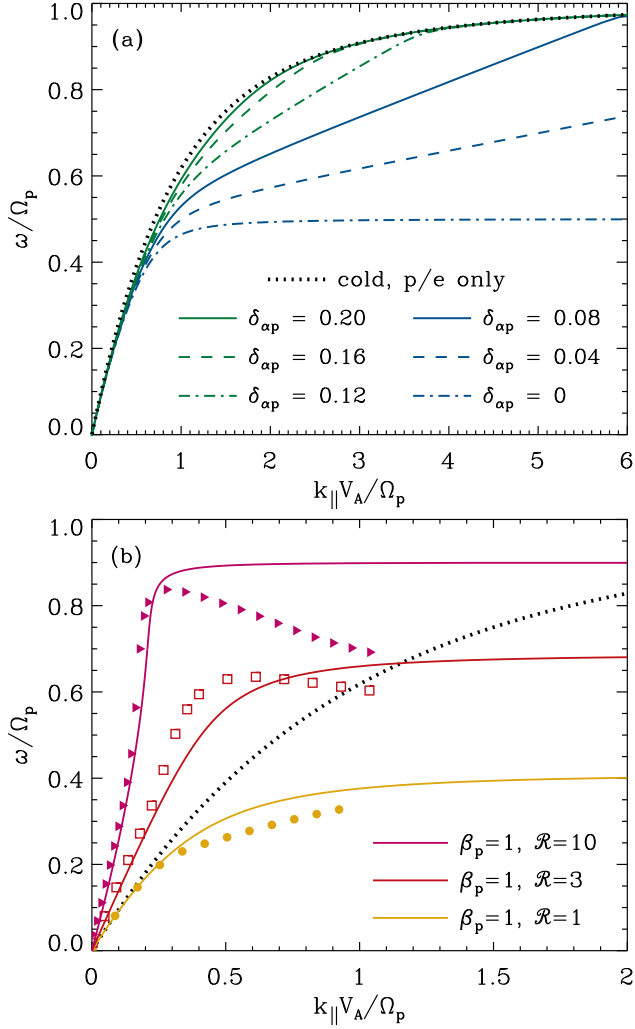}
\caption{Dispersion relations for parallel propagating ion cyclotron
waves.
In both panels, the cold plasma dispersion for a proton--electron
plasma is denoted by a black dotted curve.
(a) Variation of frequency for cold plasma with 5\% alpha particles
by number (blue and green curves, labeled by their values of
$\delta_{\alpha p}$).
(b) Warm plasma dispersion computed using Equation (\ref{eq:wad})
(solid curves) and with numerical Vlasov--Maxwell dispersion code
(symbols). These curves were computed with $\beta = 1$; see labels
for values of the anisotropy ratio ${\cal R}$.
\label{fig01}}
\end{figure}

Figure \ref{fig01}(a) shows various solutions to the cold plasma
dispersion relations.
Equation (\ref{eq:cdisp0}) describes the solution without alpha
particles, which has the highest frequency in this panel.
Figure \ref{fig01}(a) also shows the lower branch for a common
solar wind value of $h=0.05$ and a range of drift speeds
($0 < \delta_{\alpha p} < 0.2$), for which the lowest-$y$
roots of Equation (\ref{eq:cdispA}) were found numerically.
When $\delta_{\alpha p} \gtrsim 0.3$, the alpha particles have
essentially drifted out of resonance and the lower dispersion branch
is indistinguishable from the case of $h=0$ \citep[see also][]{HI02}.

An improved solution to the dispersion relation can be found by
beginning to take into account the proton thermal spread.
Including the next term in the asymptotic series expansion for
$Z(\xi)$ yields such a ``warm plasma'' dependence on
temperature.
Thus, the approximation
\begin{equation}
  Z(\xi) \, \approx \, -\frac{1}{\xi} - \frac{1}{2\xi^3}
  \label{eq:xi13}
\end{equation}
is inserted into Equation (\ref{eq:epsR}) as before.
For a pure proton--electron mixture, the analytic dispersion
relation becomes
\begin{equation}
  y^{2} = x^{2} (1-y) \left\{ 1 +
  \frac{\beta}{2 (y-1)^2} \left[ {\cal R} - 1 +
  \left( \frac{y}{y-1} \right) \right] \right\}
  \label{eq:wad}
\end{equation}
where for brevity the subscript $p$ has been removed from the
anisotropy ratio ${\cal R}$ and the parallel proton plasma beta
is defined as
\begin{equation}
  \beta \, = \, \frac{w_{p \parallel}^2}{V_{\rm A}^2}
  \, = \, \frac{8\pi n_{p} k_{\rm B} T_{p \parallel}}{B_{0}^2}
  \,\, .
\end{equation}
Equation (\ref{eq:wad}) can be solved explicitly for $x$ as a
function of $y$.

Figure \ref{fig01}(b) displays a selection of solutions to
Equation (\ref{eq:wad}) and compares them with numerical solutions
from the full Vlasov--Maxwell code of \citet{CvB03,CvB12}.
The code used in this paper is a new version that handles
bi-Maxwellian anisotropies by using the complete form of the
dispersion relation derived by \citet{Br98}.
These equations are also consistent with those of \citet{PG11}
because we limit the parameter space to $k_{\parallel} > 0$.
The numerical solutions cease to have well-behaved (i.e.,
weakly damped) solutions for $x \gtrsim 0.7 \beta^{-0.4}$
\citep[see also][]{Stix,CvB12}, but Equation (\ref{eq:wad})
provides continuous solutions for $x \rightarrow \infty$.
Despite the analytic solutions not exhibiting the local maximum
in $y(x)$ that the numerical solutions show at high values of
${\cal R}$, the overall behavior at low and intermediate values
of $x$ is captured well by Equation (\ref{eq:wad}).

The low-wavenumber limit (i.e., $|x| \ll 1$) of
Equation (\ref{eq:wad}) is
\begin{equation}
  y \, \approx \,
  x \sqrt{1 + \frac{\beta}{2} \left( {\cal R} - 1 \right)}
  \, \equiv \, \Theta x
  \label{eq:Thdef}
\end{equation}
which is the well-known version of Alfv\'{e}n wave dispersion
in the presence of anisotropic gas pressure \citep{Ba66,Is84b}.
The ideal MHD limit of $\Theta \approx 1$ occurs for
either nearly isotropic protons (${\cal R} \approx 1$) or a
very low-beta plasma.
In the high-wavenumber limit (i.e., $|x| \gg 1$), the warm
dispersion relation approaches a constant value $y_{\infty}$ that
in general has $0 < y_{\infty} < 1$.
This asymptotic frequency satisfies the cubic equation
\begin{equation}
  2 (y_{\infty}-1)^{3} + \beta {\cal R} (y_{\infty}-1) + \beta = 0
\end{equation}
and the cold limit of $\beta \rightarrow 0$ reproduces
$y_{\infty} \rightarrow 1$ as it should.
On the other hand, the ``hot'' limit of $\beta \rightarrow \infty$
is consistent with an asymptotic frequency of
$y_{\infty} \approx ({\cal R}-1)/{\cal R}$
(as long as ${\cal R} \geq 1$).
\citet{SS10} noted that there is a regime of parameter space
that does not allow for propagating solutions to the warm
Alfv\'{e}n wave dispersion relation.
For ${\cal R} < 1 - (2 / \beta)$, the parameter $\Theta$ is
imaginary and Equation (\ref{eq:wad}) has no real solutions
for the frequency.
This region of parameter space is identical to the region
described by the classical nonresonant firehose instability
threshold \citep{Ga98}.

\begin{figure}
\epsscale{1.11}
\plotone{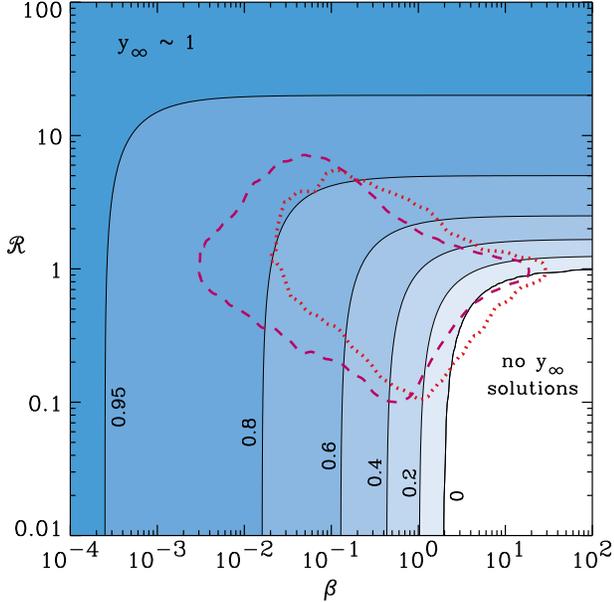}
\caption{Constant values of the asymptotic dimensionless
frequency $y_{\infty}$ shown as labeled contours vs.\  $\beta$
and ${\cal R}$.
{\em Wind} measurement outlines are shown for data presented by
\citet{Hg06} (magenta dashed curve) and
\citet{Ma12} (red dotted curve).
\label{fig02}}
\end{figure}

Figure \ref{fig02} shows how the asymptotic scaled frequency
$y_{\infty}$ varies as a function of both $\beta$ and ${\cal R}$.
The excluded firehose regime is evident on the lower right.
For heliospheric context, approximate outlines of the observed
values of $\beta$ and ${\cal R}$ at 1~AU are also plotted in
Figure \ref{fig02}.
Both curves enclose the occupied regions of parameter space
as measured by the {\em Wind} spacecraft at 1~AU, but they apply
to slightly different subsamples.
The data from \citet{Hg06} were for the slow solar wind only
($u \leq 600$ km s$^{-1}$), although their data for the fast solar
wind were fully enclosed within the slow-wind outline.
The data from \citet{Ma12} were constrained to include only
``collisional age'' parameters $A_{p} \leq 0.1$ (see
Section \ref{sec:radial:eqns} for definitions).

\subsection{Resonant Shell Distributions: Fits to Simulations}
\label{sec:disp:shell}

In a collisionless medium, the presence of cyclotron resonant waves
causes ion VDFs to evolve into distinctly non-bi-Maxwellian shapes.
\citet{KE66} and \citet{Rw66} showed that VDFs undergo diffusion
in velocity space along resonant surfaces described by kinetic
energy conservation in the wave's phase-speed reference frame.
Thus, the shapes of these resonant ``shells'' as a function of
$v_{\parallel}$ and $v_{\perp}$ depend on the details of the
dispersion relation \citep[see also][]{GS00,Is00}.
However, as demonstrated above, when one computes the dispersion
relation in anything but the cold-plasma limit, the answer
depends on the thermal spread of the particles; i.e., on the
shape of the VDF itself.
Finding a truly self-consistent solution for both the dispersion
relation and the VDF is a nontrivial problem.

For the specific case of marginal stability ($\gamma \rightarrow 0$)
in a proton--electron plasma, \citet{Is12} and \citet{Is13} found
consistent numerical solutions for both $\omega_{r}(k_{\parallel})$
and $f_{p} ( v_{\parallel}, v_{\perp} )$.
These solutions were obtained under the assumption that the density
of protons in velocity space decreases from its maximum
central value with a specified parameterization.
Nevertheless, the VDF was constrained to remain constant along the
resonant shells that were consistent with the dispersion relation.
The resulting dependence of ${\cal R}$ on $\beta$ was found to
produce better agreement
with the observed upper-right edge of the data envelopes shown in
Figure \ref{fig02} than did earlier results based on bi-Maxwellians.

Although solar wind protons are not likely to spend all of their
time right at marginal stability, it is useful
to explore how the VDF solutions of \citet{Is13} can be applied
to large-scale models of cyclotron heating in the heliosphere.
The dispersion relations shown in Figure~2 of \citet{Is13} were
reproduced by finding a parameterized fit, which we first
estimated with
\begin{equation}
  (\Theta x)^{2} \, = \, \frac{y^2}{| 1 - y |^{\phi}} \,\, ,
  \label{eq:philfit}
\end{equation}
where
\begin{equation}
  \phi \, = \, \frac{1 + 5\beta^{0.4}}{1 + 9\beta^{0.4}}
  \label{eq:phidef}
\end{equation}
and $\Theta$ is defined in Equation (\ref{eq:Thdef}).
When $\beta \rightarrow 0$, the exponent $\phi$ approaches 1 and
$\Theta \approx 1$, and thus Equation (\ref{eq:philfit}) comes
into agreement with Equation (\ref{eq:cdisp0}).
A simpler version of this equation (with $\phi = 1$
for all values of $\beta$) was used by \citet{Is84b}
to account for both warm-plasma anisotropy effects at low
wavenumber ($y \approx \Theta x$) and the cold plasma dispersion
relation's approach to $y = 1$ at high wavenumber.

\begin{figure}
\epsscale{1.11}
\plotone{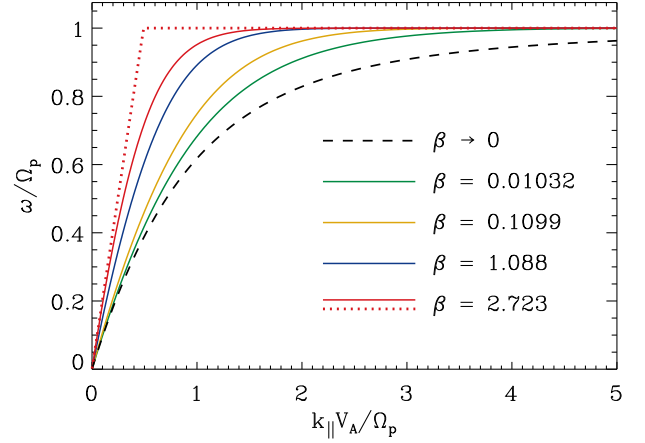}
\caption{Dispersion curves meant to reproduce Figure~2 of \citet{Is13}.
The cold plasma dispersion relation (black dashed curve) is compared
with solutions to Equations (\ref{eq:philfit})--(\ref{eq:philnew}),
with labels describing the color of each curve based on $\beta$.
Also shown is one example of the approximate hot dispersion relation
of Equation (\ref{eq:yhotqx}) (red dotted curve).
\label{fig03}}
\end{figure}

However, the dispersion relation given by
Equations (\ref{eq:philfit})--(\ref{eq:phidef}) does not accurately
replicate the results of \citet{Is13} near the cyclotron resonance
at $y \approx 1$.
Once the frequency gets close to this limiting value, the
self-consistent models were seen to approach it with exponential
rapidity.
We adjusted the solution by first solving for $y(x)$ using the
above fitting formulae and calling it $y_{\rm old}$, then we
forced the exponential behavior with
\begin{equation}
  y_{\rm new} = 1 - (1 - y_{\rm old}) e^{-\alpha |x|^3} \,\, ,
  \label{eq:philnew}
\end{equation}
where $\alpha = 0.22 \beta^{0.5}$.
The constant value of 0.22 was slightly smaller than what would be
needed to reproduce the tabulated $\alpha$ values of \citet{Is13},
but it worked best to reproduce the shape of the dispersion relation.
Figure \ref{fig03} shows the same example dispersion curves from
\citet{Is13}, but now computed with the above procedure.
The $\Theta$ parameter was computed in each case from the tabulated
pairs of $\beta$ and ${\cal R}$ listed in Table~1 of \citet{Is13}.

In the high-$\beta$ limit, the solutions shown in Figure \ref{fig03}
appear to be approaching a kind of ``hot'' dispersion relation,
which remains close to linear ($y \propto x$) until $y$ approaches
unity, then it flattens rapidly to $y=1$ for all higher wavenumbers.
The consequences of two forms of such a hot dispersion relation will
be explored further below.
The high-$\beta$ limit of our fit to the \citet{Is13} results is
described approximately by
\begin{equation}
  y \, = \, \min \left( \Theta x , 1 \right) \,\, ,
  \label{eq:yhotqx}
\end{equation}
and this is also plotted for comparison in Figure \ref{fig03} for
the case with the largest $\beta$.
We will also explore the consequences of an even simpler hot
approximation,
\begin{equation}
  y \, = \, \min \left( x , 1 \right)
  \label{eq:yhotx}
\end{equation}
which transitions from the ideal MHD dispersion relation to strict
cyclotron resonance for $x > 1$.

\subsection{Resonant Shell Distributions: Analytic Approximation}
\label{sec:disp:anshell}

It is worthwhile to investigate whether a purely analytic approach
to describing resonant shell VDFs could give rise to an improved
dispersion relation.
For the case studied below, the resulting dispersion relation
turns out to be identical to the warm bi-Maxwellian
approximation of Equation (\ref{eq:wad}).
However, we present this analysis for the sake of completeness,
and to show that some fully kinetic models may sometimes give results
not so different from those found by assuming bi-Maxwellian VDFs.

The major simplifying assumption made here is that the resonant
shells are described by contours of constant phase speed $V_{\rm A}$
for forward and backward moving Alfv\'{e}n waves.
This is consistent with the ideal MHD dispersion relation ($y=x$)
or, equivalently, Equation (\ref{eq:yhotx}) above.
Our goal is to determine to what extent the dispersion relation that
results from solving Equation (\ref{eq:epseqn}) is consistent with
that input assumption.
The proton VDFs are constant along contours described by
constant values of
\begin{equation}
  \eta^{2} \, = \, v_{\perp}^{2} +
  (V_{\rm A} + |v_{\parallel}|)^{2} - V_{\rm A}^{2}
  \label{eq:etadef}
\end{equation}
\citep[e.g.,][]{Is12},
and we use the absolute value of $v_{\parallel}$ to create a
symmetric VDF consistent with the presence of an equal population
of forward and backward moving Alfv\'{e}n waves.
The normalization used in Equation (\ref{eq:etadef}) implies that
$\eta$ is the value of $v_{\perp}$ encountered by each shell contour
when it passes through $v_{\parallel} = 0$.

Following \citet{Is13}, the VDF is defined as a Gaussian
function of the single parameter $\eta$, and the
``thermal speed'' in units of $\eta$ is specified as $\sigma$.
The bulk thermal properties of the VDF can be parameterized in
terms of a dimensionless velocity ratio $a = V_{\rm A} / \sigma$.
We normalize the distribution by requiring its zeroth moment
to be the proton number density, and
\begin{equation}
  f_{p} (\eta) \, = \,
  \frac{n_{p} \, e^{-a^2}}
  {\pi^{3/2} \sigma^{3} (1 - \mbox{erf} \, a)}
  \exp \left( - \frac{\eta^2}{\sigma^2} \right)  \,\, ,
  \label{eq:fpshell}
\end{equation}
where $\mbox{erf}$ is the error function.
In the limit of $a \rightarrow 0$, Equation (\ref{eq:fpshell})
becomes the standard isotropic Maxwellian distribution.

\begin{figure}
\epsscale{1.11}
\plotone{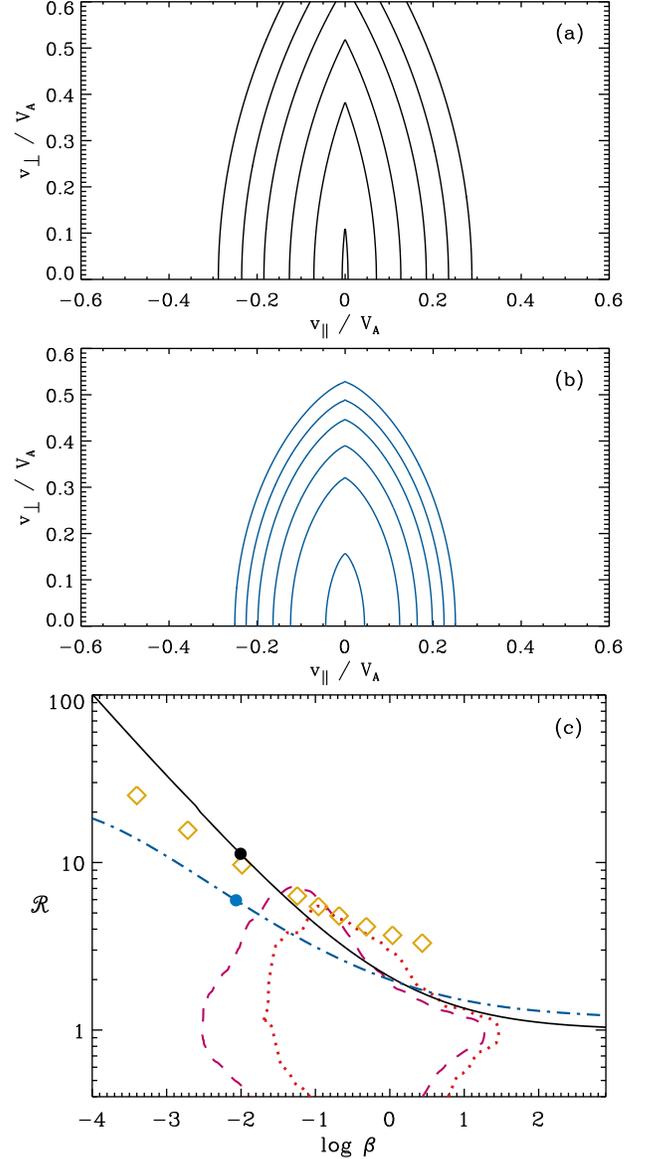}
\caption{Contours of $f_{p} ( v_{\parallel}, v_{\perp} )$ for
(a) the analytic model described by Equation (\ref{eq:fpshell})
with $a=3$, and (b) resonant shell contours consistent with
cold plasma dispersion, for $\beta = 0.01$.
(c) Loci of points in ($\beta$,${\cal R}$) space consistent
with the analytic shell model (black solid curve) and the
cold plasma model (blue dot-dashed curve), with the specific
values of $\beta$ shown in panels (a)--(b) labeled by filled
circles.
Also shown are observational outlines (same as in Figure \ref{fig02})
and the self-consistent results of \citet{Is13} (gold diamonds).
\label{fig04}}
\end{figure}

Figure \ref{fig04}(a) shows contours of the VDF described above
for a representative value of $a=3$.
The VDF exhibits ${\cal R} > 1$ anisotropy for any value of $a > 0$.
The moments of Equation (\ref{eq:fpshell}) were computed in order
to determine how $\beta$ and ${\cal R}$ depend on the single
parameter $a$.
We found that
\begin{equation}
  w_{p \perp}^{2} = \sigma^{2}
  \,\, , \,\,\,\,\,
  w_{p \parallel}^{2} = \sigma^{2} \Psi
  \,\, ,
\end{equation}
where
\begin{equation}
  \Psi = 2 a^{2} + 1 -
  \frac{2a e^{-a^2}}{\pi^{1/2} (1 - \mbox{erf} \, a)}
  \,\, ,
\end{equation}
and thus ${\cal R} = 1 / \Psi$ and $\beta = \Psi / a^{2}$.
For the case $a=3$ shown in Figure \ref{fig04}(a), the VDF
exhibits $\beta = 0.01203$ and ${\cal R} = 9.237$.
Because both ${\cal R}$ and $\beta$ are functions of a single
parameter $a$, they trace out a distinct curve in the
beta--anisotropy diagram.

Figure \ref{fig04}(b) shows a comparable set of VDF contours
that were computed to be consistent with the cold
plasma dispersion relation (Equation (\ref{eq:cdisp0})); see
Section \ref{sec:heat:diff} for more details about how this was done.
In comparison to the dispersionless VDF shown in Figure \ref{fig04}(a),
the cold plasma contours are known to be ``snubbed'' around
$v_{\parallel} \approx 0$, and thus the cold VDF exhibits less
anisotropy for similar values of $\beta$.
Figure \ref{fig04}(c) illustrates the locus of points in the
beta--anisotropy plane described by the above analytic model.
For $\beta < 1.6$, the analytic anisotropy ratio exceeds that of
the cold plasma model.
It also exceeds the self-consistent marginal anisotropy ratio
computed by \citet{Is13} for $\beta \lesssim 0.02$.

\citet{Is12} showed how the operator $\hat{\Sigma}$ can be written
in terms of a single partial derivative of $\eta$, and we applied
that expression to Equation (\ref{eq:eorig}).
For a pure proton--electron plasma with cold (Maxwellian) electrons
and $V_{\rm A} \ll c$, the dispersion relation is given by
\begin{displaymath}
  \mbox{Re} (\epsilon) \, = \, 
  \left( \frac{\omega_{pp}}{\omega_r} \right)^{2} \, \left\{
  - \frac{\omega_r}{\Omega_p}
  + \xi_{0} \left[ S_{+} (\xi_{1},a) + S_{-} (\xi_{1},a) \right]
  \right.
\end{displaymath}
\begin{equation}
  \left.
  + \, a \left[ S_{+} (\xi_{1},a) - S_{-} (\xi_{1},a) \right] \right\}
  \label{eq:eshell}
\end{equation}
where the resonance factors are defined for the present model as
\begin{equation}
  \xi_{0} \, = \, \frac{\omega_r}{k_{\parallel} \sigma}
  \,\,\, , \,\,\,\,\,\,
  \xi_{1} \, = \, \frac{\omega_{r} - \Omega_p}{k_{\parallel} \sigma}
\end{equation}
and we define the components of a generalized plasma dispersion
function as follows,
\begin{equation}
  S_{\pm} (\xi,a) \, = \,
  \frac{\pm 1}{\sqrt{\pi} (1 - \mbox{erf} \, a)}
  \int_{0}^{\pm \infty} \frac{dt \,
  e^{-(a + |t|)^2}}{t-\xi} \,\, .
  \label{eq:Spm}
\end{equation}
In the limit of $a \rightarrow 0$, the sum
$S_{+}+S_{-}$ becomes the standard Maxwellian dispersion function
$Z(\xi)$.
The difference $S_{+}-S_{-}$ does not contribute to
the dispersion relation in the $a \rightarrow 0$ limit because it
is multiplied by a factor of $a$ in Equation (\ref{eq:eshell}).

The asymptotic series expansion for large values of $|\xi|$ was
obtained for $S_{+}(\xi,a)$, and the antisymmetry property
\begin{equation}
  S_{-} (\xi, a) \, = \, -S_{+}(-\xi, a)
\end{equation}
was used to compute $S_{-}(\xi,a)$.
Unlike Equation (\ref{eq:xi13}), which has nonzero coefficients
for only the odd powers of $\xi$, the full expansion for
Equation (\ref{eq:Spm}) also has nonzero values for the even
powers, with
\begin{displaymath}
  S_{+} (\xi,a) \, = \,
  -\frac{1}{2\xi} + \frac{1}{\xi^2} \left( \frac{a}{2} - X \right)
  - \frac{1}{\xi^3} \left( \frac{2a^{2} + 1}{4} - aX \right)
\end{displaymath}
\begin{equation}
  + \, \frac{1}{\xi^4} \left[ \frac{2a^{3} + 3a}{4} - (1 + a^2) X
  \right] + \, \cdots \,\, ,
\end{equation}
where $X = e^{-a^2} / [2 \pi^{1/2} (1 - \mbox{erf} \, a)]$.
In Equation (\ref{eq:eshell}), each $1/\xi^{2n}$ (even) term ends up
having a comparable contribution to the dispersion relation as
the next higher $1/\xi^{2n+1}$ (odd) term, so we truncated the above
series expansion at $1 / \xi^{3}$.
Including those terms into Equation (\ref{eq:eshell}), the
dispersion relation was found to be
\begin{equation}
  y^{2} = x^{2} (1-y) \left[ 1 +
  \frac{y - 1 + \Psi}{2 a^{2} (y-1)^3} \right] \,\, .
  \label{eq:wadshell} 
\end{equation}
However, when the definitions of $\beta$ and ${\cal R}$ given above
are substituted in for $\Psi$ and $a$, the result is seen to be
identical to the bi-Maxwellian warm plasma relation of
Equation (\ref{eq:wad}).
Interesting as that may be, it is formally inconsistent with the
initial assumption of $y=x$ used to compute the VDF shell shapes.
This kind of analytic model deserves further study, but for now
we set it aside and follow other approaches to model the resonant
diffusion of protons in velocity space.

\section{Proton Cyclotron Heating}
\label{sec:heat}

Once the dispersion relation for parallel-propagating Alfv\'{e}n
waves has been specified, it becomes possible to estimate the rate
of energy transfer between the waves and the particles.
Section \ref{sec:heat:spec} defines the relevant wave power
quantities needed to determine how rapidly the protons are
energized, and Sections \ref{sec:heat:diff}--\ref{sec:heat:bimax}
present two different theoretical frameworks for modeling the heating.
Section \ref{sec:heat:norm} compares various estimates of the
total heating rate with one another and with observational
constraints.

\subsection{Alfv\'{e}nic Power Spectrum}
\label{sec:heat:spec}

For linear Alfv\'{e}n waves, we assume the total energy density
$U_{\rm A}$ is divided between transverse kinetic and
magnetic fluctuations.
The full three-dimensional (3D) power spectrum $E_{\rm A} ({\bf k})$
is written as a general function of vector wavenumber ${\bf k}$
and is normalized such that
\begin{equation}
  U_{\rm A} \, = \,
  \frac{1}{2} \rho_{0} \langle \delta v_{\perp}^{2} \rangle
  + \frac{\langle \delta B_{\perp}^{2} \rangle}{8\pi}
  \, = \, \int d^{3} {\bf k} \, E_{\rm A} ({\bf k}) \,\, .
  \label{eq:UA}
\end{equation}
The kinetic fluctuation strength depends on the background density
$\rho_0$ and the transverse velocity variance
$\langle \delta v_{\perp}^{2} \rangle$, and the associated magnetic
fluctuation variance is given by
$\langle \delta B_{\perp}^{2} \rangle$.
The variables defined here are similar, but not identical, to those
used by \citet{CvB03,CvB12}.

As summarized in Section \ref{sec:summ}, we assume the low-wavenumber
part of the spectrum---which contributes nearly all the power---is
the product of an ongoing MHD turbulent cascade.
This paper is not concerned much with that dominant part of the
spectrum except as a potential source of the high-$k_{\parallel}$
ion cyclotron resonant waves.
There is no agreement on the origin of the cyclotron waves, but
for now we follow \citet{Ch05} and \citet{CvB12} and assume they
arise from nonlinear mode coupling between Alfv\'{e}n waves and
compressive magnetosonic waves.
In that model, the resulting Alfv\'{e}nic power spectrum is close
to isotropic, with a $k^{-3/2}$ reduced power-law behavior
consistent with several models and simulations
\citep[e.g.,][]{Ir63,Kr65,Nk99,Bo06,Gr13}.
In order to normalize the total power to Equation (\ref{eq:UA}),
this kind of isotropic spectrum can be written as
\begin{equation}
  E_{\rm A} (k) \, = \,
  \frac{\langle \delta B_{\perp}^{2} \rangle}{32\pi^{2} k_{0}^3}
  \, \times \left\{ \begin{array}{ll}
    0 \,\, , & k < k_{0} \,\, , \\
    (k_{0}/k)^{7/2} \,\, , & k \geq k_{0} \\
  \end{array} \right.
  \label{eq:EAnodamp}
\end{equation}
where $k_0$ is a representative ``outer scale'' wavenumber.
For simplicity, we assumed equipartition between the magnetic and
kinetic fluctuations.

When working with parallel-propagating ion cyclotron waves, it is
not necessary to specify the full 3D power spectrum.
A reduced one-dimensional spectrum (i.e., a function of
$k_{\parallel}$ only) can be defined by integrating $E_{\rm A}$
over the $k_{\perp}$ coordinate.
By convention, we define the reduced power spectrum
$P_{\rm B}(k_{\parallel})$ as following only the magnetic
fluctuations, and thus it should be normalized to
\begin{equation}
  \frac{\langle \delta B_{\perp}^{2} \rangle}{8\pi}
  \, = \, \int dk_{\parallel} \, P_{\rm B} (k_{\parallel}) \,\, .
  \label{eq:PBdef}
\end{equation}
It is also consistent with our assumption of nearly
parallel-propagating waves to replace the wavenumber magnitude $k$
in Equation (\ref{eq:EAnodamp}) by $k_{\parallel}$.
Making that approximation, the reduced power spectrum can be
written as
\begin{equation}
  P_{\rm B} (k_{\parallel}) \, \approx \,
  \frac{\langle \delta B_{\perp}^{2} \rangle}{48\pi k_0}
  \, \times \left\{ \begin{array}{ll}
    0 \,\, , & k_{\parallel} < k_{0} \,\, , \\
   (k_{0}/k_{\parallel})^{3/2} \,\, , & k_{\parallel} \geq k_{0} \,\, . \\
  \end{array} \right.
  \label{eq:PBnodamp}
\end{equation}
The above expression does not apply to the low-$k$ nonresonant
part of the spectrum, so its integral over all values of
$k_{\parallel}$ does not match up with the normalization of
Equation (\ref{eq:PBdef}).
However, Equation (\ref{eq:PBnodamp}) gives the proper value of
the {\em local} reduced power (in the high-$k_{\parallel}$ regime)
in agreement with the 3D spectrum of
Equation (\ref{eq:EAnodamp}).

Another way to normalize the reduced power spectrum is to specify
its value at the nominal proton cyclotron resonant wavenumber
$k_{\parallel} = \Omega_{p} / V_{\rm A}$ (i.e., $x=1$).
Calling this normalized power level $P_1$, a simple parameterization
is given by
\begin{equation}
  P_{\rm B} (k_{\parallel}) \, = \, P_{1}
  \left( \frac{\Omega_p}{k_{\parallel} V_{\rm A}} \right)^{n}
  \, = \, \frac{P_1}{x^n}
  \,\, ,
  \label{eq:PBresnorm}
\end{equation}
where we normally assume $n = 3/2$ as above.
In typical models of the solar corona and heliosphere
\citep[e.g.,][]{CvB12}, the resonant wavenumber
$\Omega_{p} / V_{\rm A}$ is many orders of magnitude larger than
the turbulent energy-containing wavenumber $k_0$.
Although Equation (\ref{eq:PBresnorm}) is used for most of the
proton heating models described below, it is occasionally compared
with Equation (\ref{eq:PBnodamp}) in order to estimate a reasonable
range of values for $P_1$.

The above expressions assumed $k_{\parallel} > 0$ for outward
propagating waves.
However, the models below sometimes include a population of
inward propagating waves (i.e., $k_{\parallel} < 0$) as well.
When evaluating the power available for these waves, we first
specify the power in outward waves, then set the inward wave power
as a specified fraction $f_{\rm in}$ of the outward power.
Thus, the limiting case of purely outward propagating waves
corresponds to $f_{\rm in} = 0$, and the case of balanced power
between outward and inward modes (i.e., zero cross-helicity)
corresponds to $f_{\rm in} = 1$.
Details about the power spectrum for the inward waves are computed
from expressions such as Equations (\ref{eq:PBnodamp}) or
(\ref{eq:PBresnorm}), but with the absolute value of $k_{\parallel}$
or $x$ used instead of the signed quantity.
The sign of $k_{\parallel}$ matters in the resonance factors and
diffusion coefficients described below.

Lastly, we note that the power-law spectra described above do not
contain the effects of wave dissipation that are caused by the
wave-particle interactions.
These effects have been included in phenomenological models of
turbulent cascade \citep[e.g.,][]{Li01,Hw08,Ji09,CvB12}.
Section \ref{sec:alpha} treats spectral damping for the specific problem
of alpha particles sapping away the energy before the protons have
a chance to resonate with the high-$k_{\parallel}$ fluctuations.
However, for the models discussed in the remainder of this section
we continue to assume a power-law form of $P_{\rm B}(k_{\parallel})$
for simplicity.

\subsection{Quasilinear Diffusion in Velocity Space}
\label{sec:heat:diff}

The derivation of the linear dispersion relation made the assumption
that fluctuations in the VDFs and electromagnetic fields are small
first-order oscillations, and that any second- or higher-order
quantities are negligible.
To determine how the waves and particles interact with one another
to produce net heating, a so-called quasilinear approach is often
applied \citep{KE66,Rw66,GS83,MT01a}.
In quasilinear theory, second-order fluctuation quantities are
retained and averaged over spatial and time scales long in
comparison to those of the gyromotions, and random phases are
assumed for the first-order Fourier oscillations
themselves.\footnote{%
\citet{Hw06} made the case that quasilinear theory often neglects
the idea that plasma heating (i.e., an actual increase in VDF
entropy) must always involve the randomizing effect of
particle-particle collisions.
However, in low-density plasmas the heating rate can become
independent of the collision rate and thus be determined practically
by the ``collisionless'' wave-particle resonances.}

Following \citet{MT01a}, the zeroth-order VDF
$f_{s}(v_{\parallel},v_{\perp})$ for ion species $s$
evolves via diffusion in velocity space, with
\begin{displaymath}
  \frac{\partial f_s}{\partial t} \, = \,
  \frac{1}{v_{\perp}} \frac{\partial}{\partial v_{\perp}}
  \left[ v_{\perp} \left(
  D_{\perp \perp} \frac{\partial f_s}{\partial v_{\perp}} +
  D_{\perp \parallel} \frac{\partial f_s}{\partial v_{\parallel}}
  \right) \right]
\end{displaymath}
\begin{equation}
  + \, \frac{\partial}{\partial v_{\parallel}}
  \left(
  D_{\parallel \perp} \frac{\partial f_s}{\partial v_{\perp}} +
  D_{\parallel \parallel} \frac{\partial f_s}{\partial v_{\parallel}}
  \right) \,\, .
  \label{eq:dfdt}
\end{equation}
The diffusion coefficients are given by
\begin{equation}
  \left\{ \begin{array}{c}
    D_{\parallel \parallel} \\
    D_{\parallel \perp} \\
    D_{\perp \perp} \\
  \end{array} \right\} = 
  \left( \frac{2\pi q_{s}}{m_{s} c} \right)^{2} 
  \!\!\int \! dk_{\parallel} 
  \frac{P_{\rm B} (k_{\parallel})}{k_{\parallel}^2} \,
  \Gamma_{\rm res} (v_{\parallel},k_{\parallel})
  \left\{ \begin{array}{c}
    k_{\parallel}^{2} v_{\perp}^{2} \\
    k_{\parallel} v_{\perp} \Omega_{\ast} \\
    \Omega_{\ast}^{2} \\
  \end{array} \right\}
  \label{eq:difcoef}
\end{equation}
and $D_{\parallel \perp} = D_{\perp \parallel}$.
There is some disagreement in the literature about the identity
of the frequency-like variable $\Omega_{\ast}$.
\citet{Me86} and \citet{MT01a} give
$\Omega_{\ast} = \Omega_s$, but
\cite{Le71} and \citet{IV07} give
$\Omega_{\ast} = \omega_{r} - k_{\parallel} v_{\parallel}$.
These two formulations are identical to one another when the
resonance factor $\Gamma_{\rm res}$ is a Dirac delta function
(see below).
For now we continue to follow \citet{MT01a} and assume
$\Omega_{\ast} = \Omega_s$ (see also Equations 2.29--2.31
of \citet{KE66}), but in future work we will explore whether the
use of the other definition produces qualitatively different results.

In the standard weak-damping limit of quasilinear theory, the
cyclotron resonance factor $\Gamma_{\rm res}$ is defined as
\begin{equation}
  \Gamma_{\rm res} (v_{\parallel},k_{\parallel}) \, = \,
  \delta ( \omega_{r} - k_{\parallel} v_{\parallel} - \Omega_{s} )
  \label{eq:Gdelta}
\end{equation}
and applying this Dirac delta function to Equation (\ref{eq:difcoef})
transforms the integration over $k_{\parallel}$ into a trivial
selection of a single resonant wavenumber.
For parallel propagating waves obeying a single-branch dispersion
relation with $\omega_{r} \lesssim \Omega_{p}$, it is usually the
case that $v_{\parallel} < 0$ is required for proton resonance with
outward propagating waves and $v_{\parallel} > 0$ is needed for
resonance with inward propagating waves.
When those conditions do not apply, the argument of the delta
function is never zero no matter the value of $k_{\parallel}$, so
there is thought to be no diffusion in those parts of velocity space.

It is sometimes overlooked that Equation (\ref{eq:Gdelta}) is an
approximate limiting case of a more general resonance factor that
applies for arbitrary values of $\gamma$, the imaginary part of
the frequency.  The general version is given by
\begin{equation}
  \Gamma_{\rm res} (v_{\parallel},k_{\parallel}) \, = \,
  \frac{| \gamma / \pi |}{\gamma^{2} +
  ( \omega_{r} - k_{\parallel} v_{\parallel} - \Omega_{s} )^{2}}
  \label{eq:Glorentz}
\end{equation}
and it tends toward the limit of a Dirac delta function as
$\gamma \rightarrow 0$.
\citet{GS03} noted that particle-in-cell simulations of proton
cyclotron diffusion exhibit a smearing effect in $v_{\parallel}$
that could be due to the fact that $\gamma \neq 0$.
Nonresonant regions of velocity space that exhibit no nonzero
solutions to Equation (\ref{eq:Gdelta}) instead exhibit a small---but
not negligible---diffusion coefficient due to the more spread-out
nature of Equation (\ref{eq:Glorentz}).

In order to evaluate Equation (\ref{eq:Glorentz}), we estimated
$\gamma (k_{\parallel})$ by using a weak-damping approximation that
is often applied in tandem with quasilinear theory,
\begin{equation}
  \frac{\gamma}{\omega_r} \, = \,
  -\frac{\mbox{Im} (\epsilon)}{2 \,\, \mbox{Re} (\epsilon)}
  \label{eq:gow}
\end{equation}
with
$\mbox{Re}(\epsilon) = k_{\parallel}^{2} c^{2} / \omega_{r}^{2}$ and
\begin{equation}
  \mbox{Im} (\epsilon) \, = \,
  -\frac{4\pi^3}{\omega_{r}^{2} k_{\parallel}}
  \sum_{s} \frac{q_{s}^2}{m_s} \int dv_{\perp} \, v_{\perp}^{2}
  \left( \Omega_{s} \frac{\partial f}{\partial v_{\perp}}
  + k_{\parallel} v_{\perp} \frac{\partial f}{\partial v_{\parallel}}
  \right)_{\rm res}  \,\, .
\end{equation}
The subscript ``res'' constrains the quantity in parentheses
to be evaluated at the value of $v_{\parallel}$ that satisfies the
resonance condition
$\omega_{r} - k_{\parallel} v_{\parallel} - \Omega_{s} = 0$.
This is not fully self-consistent, since it implicitly uses the
delta function assumption of Equation (\ref{eq:Gdelta}), but it
represents an iterative step toward an improved solution.
When computing $\gamma$, we used only the proton contribution to
the sum over particle species $s$.
This component of the damping rate is often written as $\gamma_p$.

\begin{figure}
\epsscale{1.15}
\plotone{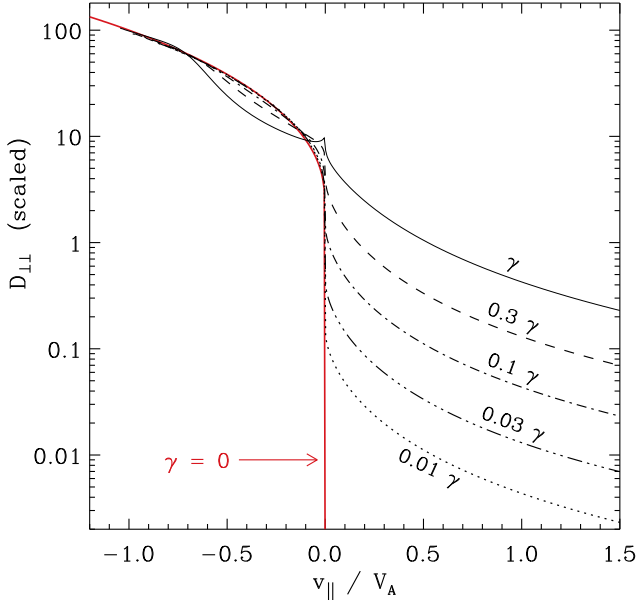}
\caption{Scaled diffusion coefficient $D_{\perp\perp}$ plotted
vs.\  parallel velocity $v_{\parallel}$, computed with various
approximations for the damping rate $\gamma$.
The result of Equation (\ref{eq:Gdelta}) (red solid curve)
is compared with that of Equation (\ref{eq:Glorentz}) for
both the actual estimated value of $\gamma(k_{\parallel})$
(black solid curve) and a range of artificially reduced damping
rates (see curve labels for reduction factors).
\label{fig05}}
\end{figure}

Figure \ref{fig05} shows example calculations of the magnitude of
the diffusion coefficient $D_{\perp\perp}$ as a function of
$v_{\parallel}$ (here computed for $v_{\perp} = 0$).
These curves were computed for purely outgoing waves
($f_{\rm in}=0$) obeying the cold plasma dispersion relation and an
isotropic Maxwellian proton VDF with $\beta = 0.1$
As expected, the idealized delta function resonance factor gives
rise to finite values of $D_{\perp\perp}$ only for
$v_{\parallel} < 0$.
However, the more accurate version (Equation (\ref{eq:Glorentz}))
produces nonzero values of $D_{\perp\perp}$ for all values of
$v_{\parallel}$.
Figure \ref{fig05} also shows how the diffusion coefficient approaches
the appropriate $\gamma \rightarrow 0$ limit when the damping rate
is multiplied by a range of arbitrary reduction factors.

In Figure \ref{fig05} there is a small cusp of increased diffusivity
around $v_{\parallel} = 0$ for the model with no reduction in $\gamma$.
The behavior of $D_{\perp\perp}$ at this velocity depends on
the high-$k_{\parallel}$ (i.e., $x \gg 1$) limiting behavior of
both the dispersion relation and the power spectrum $P_{\rm B}$.
This calculation was done with the power-law version of $P_{\rm B}$
described by Equation (\ref{eq:PBnodamp}).
Presumably, if the damping implied by our computed value
of $\gamma$ was applied self-consistently to the high-wavenumber
part of the power spectrum, the appearance of this cusp would be
significantly muted.

The diffusion coefficients describe the shapes of resonant
shell contours in velocity space toward which the VDF should evolve
as $t \rightarrow \infty$ in Equation (\ref{eq:dfdt}).
At any given value of $v_{\parallel}$ and $v_{\perp}$, one can
estimate the local angle $\alpha$ between the shell contour and the
$v_{\parallel}$ axis as
\begin{equation}
  \tan\alpha \, \approx \,
  \frac{D_{\parallel\perp}}{D_{\parallel\parallel}} \,\, .
  \label{eq:tana}
\end{equation}
Figure \ref{fig04}(b) shows the result of tracing out these contours
for an example proton VDF with $\beta = 0.01$ and waves obeying the
cold plasma dispersion relation.
This calculation assumed Equation (\ref{eq:Gdelta}) for
outward propagating waves resonant with $v_{\parallel} < 0$ protons,
and the shell shapes were reflected around $v_{\parallel} = 0$ to
show the contours for inward propagating waves resonant with
$v_{\parallel} > 0$ protons.
As described above, the cold plasma dispersion relation gives rise to
marginally stable proton VDFs that are more isotropic than if the
shell shapes were computed with ideal MHD dispersion (shown in
Figure \ref{fig04}(a)).

Equation (\ref{eq:dfdt}) is solved numerically with a similar
explicit finite differencing technique as that of \citet{Cr01}.
The standard benchmark case discussed below is a pure proton--electron
plasma with waves obeying the cold dispersion relation.
The initial condition is always a bi-Maxwellian proton VDF.
The numerical diffusion code recomputes $\gamma (k_{\parallel}$)
at each time step using Equation (\ref{eq:gow}).
The code runs slowly when using Equation (\ref{eq:Glorentz})
for every calculation of $\Gamma_{\rm res}$, since in this case the
full integration over $k_{\parallel}$ must be performed numerically.
Thus, the results shown below were obtained by using
Equation (\ref{eq:Gdelta}) in resonant regions of velocity space
(i.e., where there exist values of $k_{\parallel}$ that satisfy the
proton resonance condition
$\omega_{r} - k_{\parallel} v_{\parallel} - \Omega_{p} = 0$)
and Equation (\ref{eq:Glorentz}) elsewhere.
In order that the solutions for $\Gamma_{\rm res}$ be continuous as
a function of $v_{\parallel}$, we also used
Equation (\ref{eq:Glorentz}) in resonant regions of velocity space
with $|v_{\parallel}| \leq w_{p \parallel}$.

\begin{figure*}
\epsscale{1.13}
\plotone{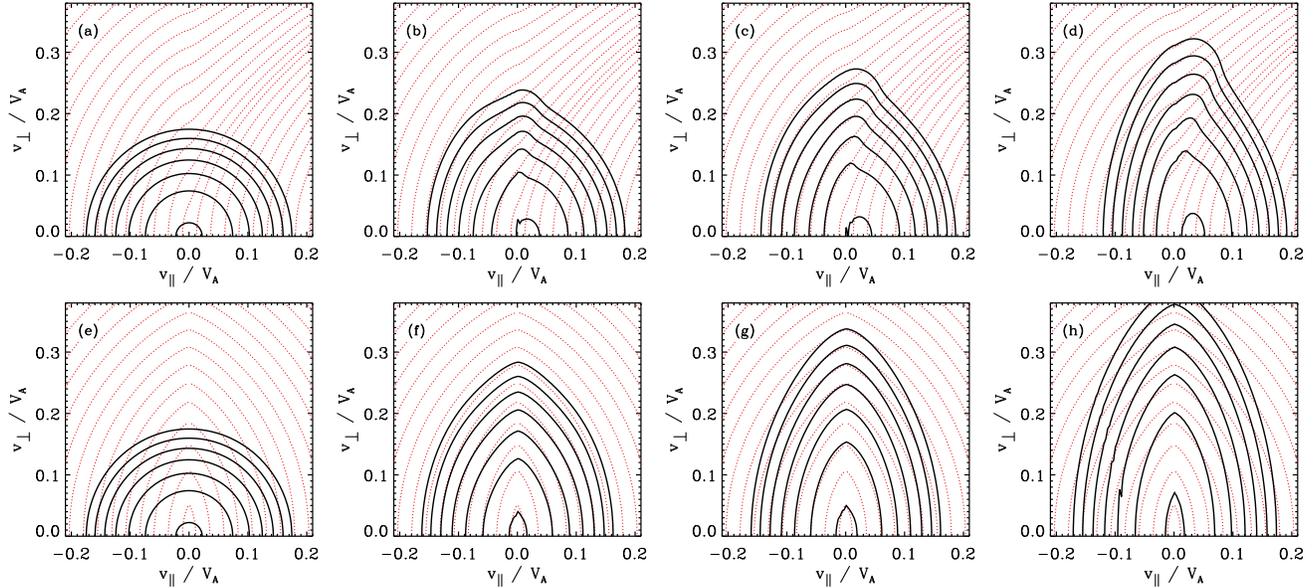}
\caption{Contour plots of the proton VDF (black curves) shown at
four times in its evolution.
From left to right, $t = 0$, 0.24, 0.48, and 2.4 in units of $\tau_d$.
The models assumed $f_{\rm in}=0$ (top row) and $f_{\rm in}=1$
(bottom row).
VDF contours are separated by constant factors of 0.5 in $\log f_{p}$.
Red dotted contours show sets of resonant shells computed by tracing
streamlines using Equation (\ref{eq:tana}).
For clarity, in the bottom row we show only the ``dominant'' set of
resonant shells in each half of velocity space.
\label{fig06}}
\end{figure*}

Figure \ref{fig06} illustrates the time evolution of proton VDFs
computed by the numerical diffusion code.
Two runs are shown, both with initial conditions of $\beta = 0.01$
and ${\cal R} = 1$.
The upper set of panels shows the evolution with $f_{\rm in} = 0$
(all outward wave power), and the lower set shows the result of
assuming $f_{\rm in} = 1$ (balanced outward and inward power).
The evolution time is specified in units of a characteristic
timescale $\tau_d$ for perpendicular diffusion, with
\begin{equation}
  \tau_{d} \, = \, w_{p \perp}^{2} / D_{\perp \perp} \,\, ,
\end{equation}
where $D_{\perp \perp}$ is evaluated at the peak of the initial VDF
($v_{\parallel} = v_{\perp} = 0$) at $t=0$.
The explicit nature of the finite differencing technique necessitated
the use of a small time step of order $10^{-3} \tau_{d}$.

The VDFs shown in Figure \ref{fig06} initially approach the
marginally stable shell contours (red dotted curves) that we
estimated from Equation (\ref{eq:tana}), but they appear later to
diffuse into more perpendicularly anisotropic shapes.
The VDF of the model with $f_{\rm in}=0$ resembles the numerical
results of \citet{GS03}.
In nonresonant parts of velocity space, the quasi-resonant shell
contours that we found by tracing the $\alpha$ angle have roughly
hyperbolic shapes.
This causes the initially isotropic VDF in the $v_{\parallel} > 0$
region to diffuse into the perpendicular direction---despite the
absence of a classical resonance condition there---and for the peak
of the VDF to migrate to a slightly higher value of $v_{\parallel}$
as well.

The model with balanced inward and outward wave power (lower
panels of Figure \ref{fig06}) undergoes substantial additional
diffusion because of the existence of resonant shells that cross
over one another.
\citet{Is01} suggested this could give rise to augmented
perpendicular heating akin to the stochastic energization found
in second-order Fermi acceleration.
More surprisingly, our model with $f_{\rm in}=0$ (upper panels)
appears to undergo extra diffusion of this type as well.
This is likely to be the result of the ``spreading'' inherent in
Equation (\ref{eq:Glorentz}), such that the resonant contours
represent merely the centroids of a range of possible diffusion
pathways in velocity space.

\begin{figure}
\epsscale{1.10}
\plotone{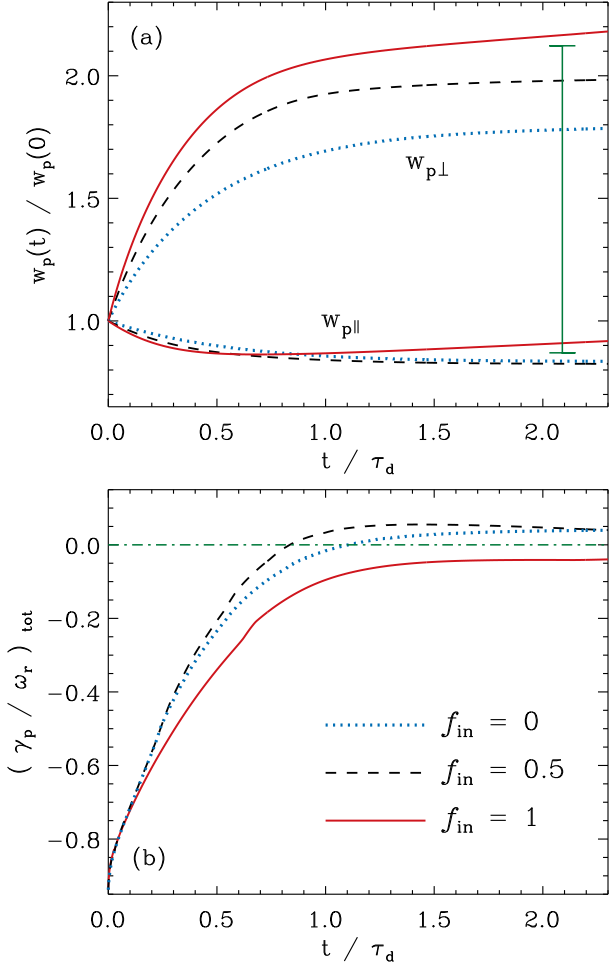}
\caption{Time evolution of (a) thermal speed moments $w_{p \perp}$
(upper curves) and $w_{p \parallel}$ (lower curves) normalized
to their initial values, and (b) the wavenumber integrated damping
rate defined by Equation (\ref{eq:gamtot}).
In both panels, the models correspond to
$f_{\rm in}=0$ (blue dotted curve),
$f_{\rm in}=0.5$ (black dashed curve), and
$f_{\rm in}=1$ (red solid curve).
In the upper panel, the green strut illustrates the magnitude
of the ``filled-shell'' anisotropy ratio from the corresponding
model shown in Figure \ref{fig04}(b).
\label{fig07}}
\end{figure}

Additional information about the quasilinear diffusion in these
models can be seen by plotting the time dependence of the VDF
moments $w_{p \parallel}$ and $w_{p \perp}$.
\citet{Ho99} discussed the conditions for net perpendicular heating
and parallel cooling of protons in resonance with cyclotron waves.
Figure \ref{fig07}(a) illustrates this evolution for the two models
discussed above and an intermediate model with $f_{\rm in}=0.5$.
The initial rates of heating and cooling are faster for the
models with inward propagating waves, since the total power
present in the system is proportional to $1 + f_{\rm in}$.
Thus, it is not surprising that the model with $f_{\rm in}=1$
has roughly twice as steep an initial slope as the model with
$f_{\rm in}=0$.
The models with additional inward power not only evolve more
rapidly, but they also begin to approach larger asymptotic values
of $w_{p \perp}$ and $w_{p \parallel}$ because of the Fermi-like
effect discussed above.

Figure \ref{fig07}(b) shows the time evolution of a
wavenumber-integrated damping rate
\begin{equation}
  \left( \frac{\gamma}{\omega_r} \right)_{\rm tot} \, = \,
  \frac{V_{\rm A}}{\Omega_p} \int dk_{\parallel} \,\,
  \frac{\gamma (k_{\parallel})}{\omega_{r} (k_{\parallel})}
  \label{eq:gamtot}
\end{equation}
\citep[see also][]{Cr01}.
The isotropic initial condition undergoes substantial wave
damping consistent with the rapid evolution to higher anisotropy.
The $f_{\rm in}=0$ model approaches an asymptotic steady state with
a positive (unstable) value of $(\gamma / \omega_{r})_{\rm tot}$
because its $v_{\parallel}>0$ shells never become completely
``filled'' in a marginally stable way.
On the other hand, the $f_{\rm in}=1$ model remains stable for
the entire simulation and evolves monotonically toward an
asymptotic state with net damping.
It is interesting that the subsequent Fermi-like diffusion away
from the resonant shells shown in Figure \ref{fig06} is still
consistent with a stable late-time evolution with $\gamma < 0$.
The model with $f_{\rm in} = 0.5$ first becomes even more unstable
than the other two models, but eventually the Fermi-like
diffusion occurs and drives $(\gamma / \omega_{r})_{\rm tot}$
slowly back down to $\gamma \approx 0$.

The numerical diffusion models shown above followed the proton VDFs
from their initial state in ($\beta$,${\cal R}$) space to an
asymptotic final state near the marginal stability curve.
However, a truly self-consistent model should have recalculated
the dispersion relation (which depends on the evolving VDF shape)
at each time step.
In lieu of tackling this extremely complex problem
\citep[see, e.g.,][]{Is12,Is13}, we now limit ourselves to measuring
only the {\em initial rates of change} away from the VDF at $t=0$.
This may be a more practical way of studying how the system
evolves when it is far from the marginal stability curve.
Following earlier work such as \citet{Ar76}, the rates of
net heating or cooling are defined as
\begin{equation}
  \left\{ \begin{array}{c}
    Q_{p \parallel} \\
    Q_{p \perp} \\
  \end{array} \right\} \, = \,
  n_{p} k_{\rm B} \frac{\partial}{\partial t}
  \left\{ \begin{array}{c}
    T_{p \parallel} / 2 \\
    T_{p \perp} \\
  \end{array} \right\}
  \label{eq:Qdef}
\end{equation}
such that their sum is the time rate of change of the total proton
internal energy density ($3 n_{p} k_{\rm B} T_{p} / 2$).
The one-fluid proton temperature is defined as
$T_{p} = (T_{p \parallel} + 2 T_{p \perp})/3$, and the partial
time derivatives are assumed to apply only for early times
$t \ll \tau_{d}$.

\begin{figure*}
\epsscale{0.93}
\plotone{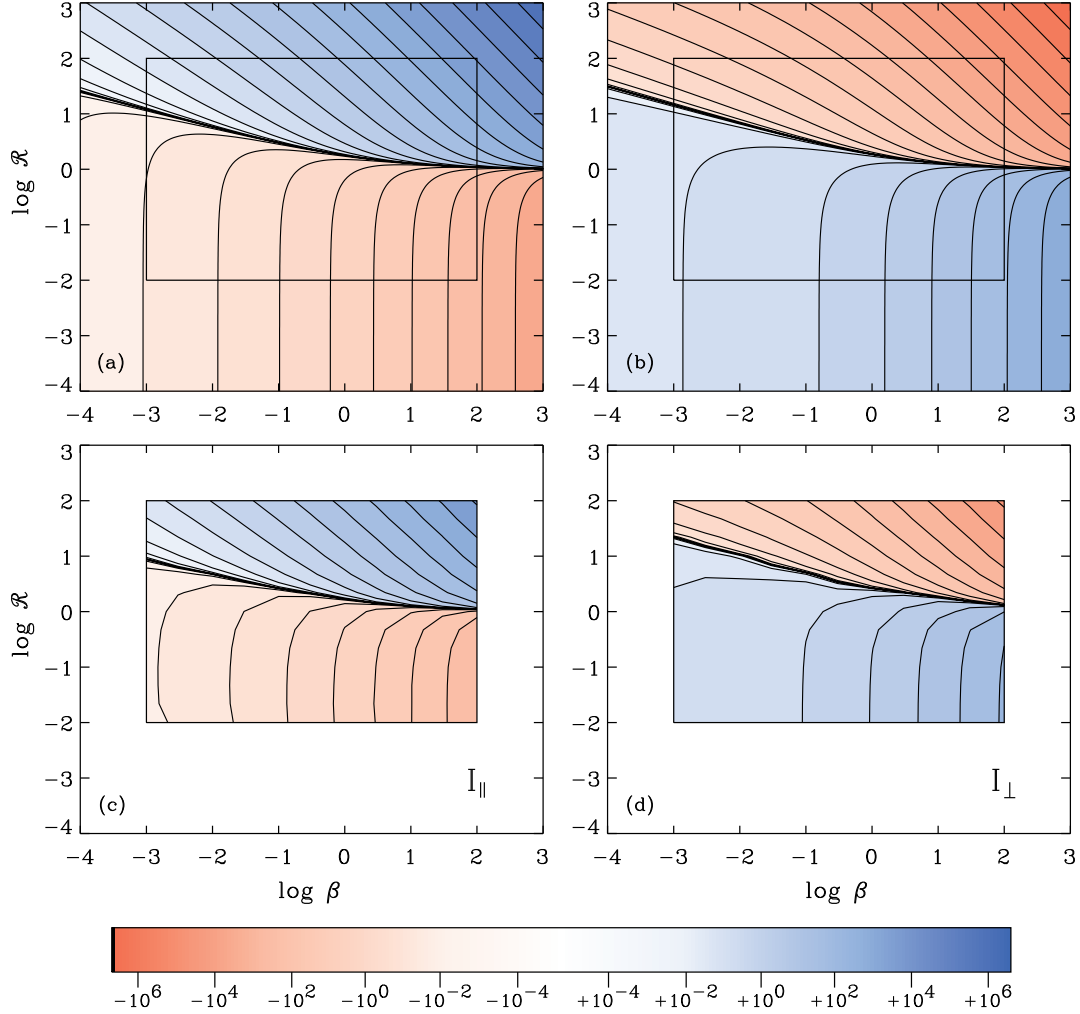}
\caption{Contours of the signed, dimensionless heating rates
$I_{p \parallel}$ (left panels) and $I_{p \perp}$ (right panels).
Parameter values are given by the color bar at bottom, and the
solid black contours are separated from their neighbors by
multiplicative factors of 4.
Panels (a)--(b) show bi-Maxwellian solutions of
Equation (\ref{eq:biheat}) and panels (c)--(d) show numerical
VDF diffusion results from Equation (\ref{eq:Qdef}).
The square box in the upper panels highlights the subset of
parameter space spanned by the models in the lower panels.
\label{fig08}}
\end{figure*}

Figure \ref{fig08} shows contours of the early-time heating rates
computed for a range of initial bi-Maxwellian VDFs.
The upper panels show the fully bi-Maxwellian approximation
described in Section \ref{sec:heat:bimax}, and the lower panels
show a coarser (11 $\times$ 13) grid of results of the
numerical diffusion code discussed above.
The plotted quantity is a dimensionless version of the
heating/cooling rate,
\begin{equation}
  I_{p \parallel, \perp} \, = \,
  \frac{V_{\rm A} \, Q_{p \parallel, \perp}}{4 P_{1} \Omega_{p}^2}
  \,\, .
  \label{eq:Idef}
\end{equation}
For the numerical diffusion models, a cold plasma dispersion relation
with pure outward waves ($f_{\rm in}=0$) was assumed.
The heating rates were computed by fitting the evolution of
$w_{p \parallel}^{2}(t)$ and $w_{p \perp}^{2}(t)$ with linear slopes.
Each case was run for 50 time steps, where one time step was given
roughly by $10^{-5} \tau_{d}$.
At these early times, there were no significant deviations from linear
increases/decreases in $w_{p \parallel}^{2}$ and $w_{p \perp}^{2}$.

Note from Figures \ref{fig08}(c)--(d) that the two-dimensional
($\beta$,${\cal R}$) space is divided by the so-called marginal
stability curve.
``Below'' that curve (i.e., for the lowest values of ${\cal R}$,
including ${\cal R}=1$), $I_{p \parallel}$ is negative and
$I_{p \perp}$ is positive.
The opposite is the case above the curve.
Thus, the net effect of cyclotron heating is to drive a plasma
either up from below or down from above, in this diagram, to
approach marginal stability as an asymptotic final state.
The curve that defines $I_{p \parallel}=0$ is not identical to
the curve that defines $I_{p \perp}=0$.
However, these two loci remain sufficiently close to one another
that they can be described more or less as a single curve.
Their degree of relative separation depends slightly on the
properties of the dispersion relation (see below).

\subsection{Bi-Maxwellian Heating and Cooling Rates}
\label{sec:heat:bimax}

If the proton VDFs are assumed to always have a bi-Maxwellian
shape as described by Equation (\ref{eq:fbimax}), the
velocity-space diffusion coefficients and the derivatives in
Equation (\ref{eq:dfdt}) can be evaluated explicitly as functions
of $T_{p \parallel}$ and $T_{p \perp}$ \citep[e.g.,][]{MT01a}.
With this assumption, the heating and cooling rates can be written
\begin{equation}
  \left\{ \begin{array}{c}
    Q_{p \parallel} \\
    Q_{p \perp} \\
  \end{array} \right\} \, = \,
  -4 \int dk_{\parallel} \, P_{\rm B} (k_{\parallel}) \,
  \frac{\gamma_p}{\omega_r}
  \left\{ \begin{array}{c}
    \omega_{r} - \Omega_{p} \\
    \Omega_{p} \\
  \end{array} \right\} \,\, ,
  \label{eq:biheat}
\end{equation}
where the bi-Maxwellian version of the proton damping rate is
given by
\begin{equation}
  \frac{\gamma_p}{\omega_r} \, = \,
  -\frac{\pi^{1/2} e^{-\xi_{1}^2}}{2} 
  \left( \frac{\Omega_p}{k_{\parallel} V_{\rm A}} \right)^{2}
  \left( {\cal R} \xi_{1} +
  \frac{\Omega_p}{k_{\parallel} w_{p \parallel}} \right) \,\, ,
  \label{eq:bigamma}
\end{equation}
and $\xi_{1}$ is the resonance factor defined in Equation (\ref{eq:xin}).
A dimensionless form of the heating rates is given by
\begin{equation}
  \left\{ \begin{array}{c}
    I_{p \parallel} \\
    I_{p \perp} \\
  \end{array} \right\} \, = \,
  - \int \frac{dx}{x^n} \left( \frac{\gamma_p}{\omega_r} \right)
  \left\{ \begin{array}{c}
    y - 1 \\
    1 \\
  \end{array} \right\} \,\, ,
  \label{eq:biI}
\end{equation}
where $x$ and $y$ are the scaled wavenumber and frequency
variables defined in Equation (\ref{eq:xydef}),
and $n$ is the power spectrum exponent that we tend to fix at 3/2.

Figures \ref{fig08}(a)--(b) show the result of solving
Equation (\ref{eq:biI}) on a fine two-dimensional grid of
$\beta$ and $\cal R$ values.
As above, we assumed a cold plasma dispersion relation with
$f_{\rm in}=0$.
For each point in the grid, the wavenumber integral over $x$
was computed on a logarithmic scale from $x = 10^{-3}$ to
$10^{+3}$.
The smallest values of $x$ tend not to contribute to the integral
because $\gamma_p$ grows exponentially small for $x \ll 1$,
and the largest values of $x$ do not contribute because of the
power spectrum falloff of $x^{-n}$.
A comparison of the upper and lower panels of Figure \ref{fig08}
shows that the bi-Maxwellian heating rates are always quite similar
in magnitude to the heating rates computed from the numerical VDF
diffusion model.
This may not be surprising, since the numerical models above were
computed only for early times when the VDF presumably remains
close to its initial bi-Maxwellian shape.
Nevertheless, these early-time rates may be the most appropriate
ones to use when studying how the plasma state evolves across wide
swaths of the ($\beta$,${\cal R}$) diagram.
The remainder of this paper will assume bi-Maxwellian VDFs for
computing $I_{p \parallel}$ and $I_{p \perp}$.

The models shown in Figure \ref{fig08} were computed assuming
outward waves only (i.e., $f_{\rm in}=0$).
However, the appearance of these contours would not be all that
different had other values of $f_{\rm in}$ been utilized (as long
as the total wave power was normalized in a consistent way).
Thus, in the bi-Maxwellian models described below we will use
$f_{\rm in}=0$ and assume that the effects of inward resonances
can be taken into account by increasing the wave power
quantity $P_1$.

For a given set of model assumptions, the marginal stability curve
can be estimated to be the locus of points where either
$Q_{p \parallel}=0$ or $Q_{p \perp}=0$.
One can also define a curve on which there is a zero rate of
change in the anisotropy ratio ${\cal R}$.
For the model of cyclotron heating studied here, this latter
curve always falls in between the two (already closely spaced)
curves that denote $Q_{p \parallel}=0$ and $Q_{p \perp}=0$.
Thus, we choose to use this condition, with
\begin{equation}
  \frac{\partial {\cal R}}{\partial t} \, = \,
  \frac{Q_{p \perp} - 2 {\cal R} Q_{p \parallel}}
  {n_{p} k_{\rm B} T_{\parallel}} \, = \, 0 \,\, ,
\end{equation}
as a practical concordance definition of marginal stability
for the bi-Maxwellian heating model.

\begin{figure*}
\epsscale{0.93}
\plotone{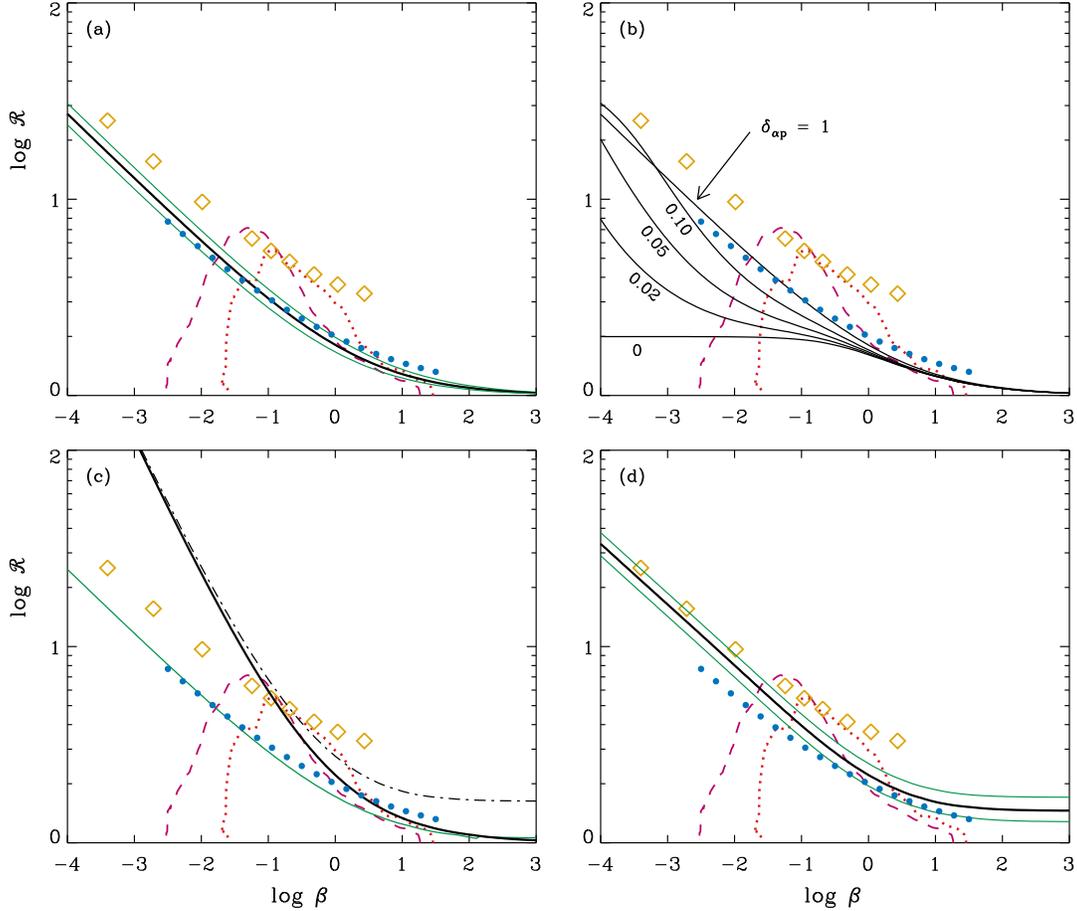}
\caption{Theoretical marginal stability curves compared in each
panel with measurements at 1~AU (red and magenta curves; see
Figure \ref{fig02}), numerical results of \citet{Is13}
(gold diamonds), and the parallel cyclotron instability threshold
of \citet{Ma12} (blue filled circles).
Specific model parameters for the four panels (black and green
curves) are enumerated in the text.
\label{fig09}}
\end{figure*}

Figure \ref{fig09} shows marginal stability curves
defined by $\partial {\cal R} / \partial t = 0$ for a range of
different dispersion relations, power spectrum indices, and
alpha--proton relative velocities.
Each of these curves was extracted from a full grid of heating rates
$I_{p \parallel}$ and $I_{p \perp}$,
similar to the ones shown in Figure \ref{fig08}(a)--(b).
These different cases also exhibit modest differences in the
magnitudes of the heating rates {\em away} from marginal stability;
this will be explored further in Section \ref{sec:radial:result1}.
The specific models included in the four panels of Figure \ref{fig09}
are described below.
\begin{enumerate}
\item
Figure \ref{fig09}(a) shows marginal stability curves computed with
a cold plasma dispersion relation, no alpha particles, and a range
of power spectrum exponents $n = 0.5$, 1.5, and 3.
Larger values of $n$ correspond to lower threshold values of
${\cal R}$ for marginal stability, but the curves do not move up
or down by very much from the standard intermediate case of $n=1.5$
(thick black curve).
These models correspond rather closely to the marginal stability
curve of \citet{Ma12} that was computed using a numerical
Vlasov--Maxwell dispersion code with bi-Maxwellian proton VDFs
(blue symbols).
\item
Figure \ref{fig09}(b) shows the result of using the cold proton--alpha
dispersion relation given by Equation (\ref{eq:cdispA}), with
$n=1.5$, $h=0.05$, and a range of relative drift speeds
$\delta_{\alpha p}$.
The largest value of $\delta_{\alpha p} = 1$ corresponds to the
alpha particles being fully out of resonance.
Thus, its curve is indistinguishable from the corresponding curve
in Figure \ref{fig09}(a).
When $\delta_{\alpha p}$ approaches zero, the marginal stability
curve moves down to substantially lower values of ${\cal R}$.
This occurs because the smaller frequency on the lower proton--alpha
dispersion branch (Figure \ref{fig01}(a)) makes a major change in
the boundary between regions of positive and negative
$I_{p \parallel}$ (see Equation (\ref{eq:biI})).
\item
Figure \ref{fig09}(c) explores the result of utilizing the warm
and hot dispersion relations derived in
Sections \ref{sec:disp:bimax}--\ref{sec:disp:shell}.
From bottom to top, the curves show the bi-Maxwellian warm
dispersion relation (Equation (\ref{eq:wad}); green solid curve),
the simplest version of the hot dispersion relation
(Equation (\ref{eq:yhotx}); thick black curve),
and the hot dispersion relation with anisotropic pressure
(Equation (\ref{eq:yhotqx}); black dot-dashed curve).
These calculations assumed $h=0$ and $n=1.5$.
The warm dispersion curve is nearly identical to that computed with
the cold plasma dispersion relation, and also with the numerical
results of \citet{Ma12}.
However, the curves computed with the hot dispersion relations
extend to substantially higher values of ${\cal R}$ than are seen
in other models.
For $-1.3 \lesssim \log\beta \lesssim -0.6$, the hot curves agree
well with the upper edge of the measured range of anisotropy ratios.
\item
Figure \ref{fig09}(d) shows marginal stability curves computed
from our fits to the \citet{Is13} dispersion relations, as
described by Equations (\ref{eq:philfit})--(\ref{eq:philnew}).
The three curves show a range of power spectrum exponents
$n = 0.5$, 1.5, and 3, with larger values of $n$ corresponding
to lower values of ${\cal R}$ as in Figure \ref{fig09}(a).
At low values of $\beta$, these curves closely approach the
numerical results of \citet{Is13} (gold diamonds).
At high values of $\beta$, the models disagree with the numerical
results but still approach an asymptotic value of ${\cal R} > 1$
similar to the hot anisotropic model shown in Figure \ref{fig09}(c).
Both of those models have dispersion relations that reduce to
$y \approx \Theta x$ in the limit of high $\beta$ and low
$k_{\parallel}$.
\end{enumerate}

\noindent
The diversity of curve shapes in Figure \ref{fig09} is somewhat
surprising, since all of these results were computed from dispersion
relations built on either cold plasma or bi-Maxwellian foundations.
None of them agree exactly with marginal stability curves
computed from shell-shaped proton VDFs, like the numerical
results of \citet{Is13} or the analytic shell model of
Section \ref{sec:disp:anshell} (see the black curve in
Figure \ref{fig04}(c)).
In a truly self-consistent model, the position of the marginal
stability curve in the beta--anisotropy plane must evolve in time
as the VDFs evolve in shape.

\subsection{Total Heating Rate Comparisons}
\label{sec:heat:norm}

\begin{figure}
\epsscale{1.15}
\plotone{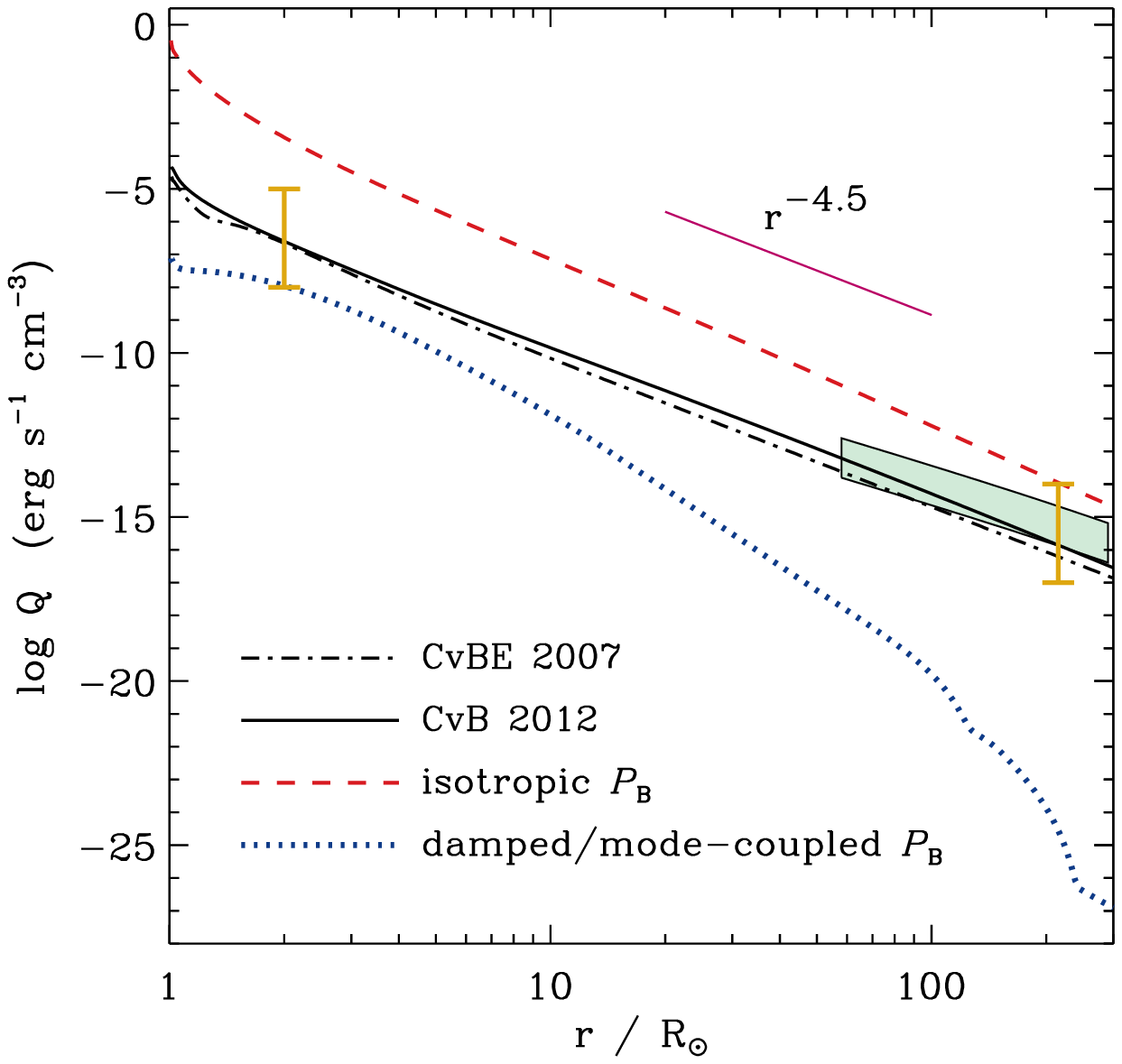}
\caption{Radial dependence of heating rates for various models
of the fast wind.
Total ($Q_{p}+Q_{e}$) rates from turbulence models of
\citet{CvB07} (black dot-dashed curve) and
\citet{CvB12} (black solid curve) agree with rates
inferred from in~situ measurements \citep{Cm09} (green region).
Proton cyclotron heating rates $Q_p$ are also shown for an
assumed isotropic wave spectrum (red dashed curve) and the
damped anisotropic spectrum of \citet{CvB12} (blue dotted curve).
Ranges of imposed heating rates in the models of
Section \ref{sec:radial:result2} are shown with gold bars.
\label{fig10}}
\end{figure}

Prior to applying the above description of proton cyclotron resonance
to a model of the solar wind, the absolute normalization for the
heating rate (which depends on the wave power spectrum) must be
specified.
As discussed in Section \ref{sec:summ}, the {\em total} rate of
plasma heating may be the result of several different physical
processes.
Thus, here we aim to explore the range of likely values for both
the total heating rate and the contribution from cyclotron
resonance.
Figure \ref{fig10} shows $Q_{p} = Q_{p \parallel}+Q_{p \perp}$
versus heliocentric distance $r$ for several assumptions about
the heating.
Black curves indicate the total heating rates (protons plus electrons)
from the fast-wind turbulence models of \citet{CvB07} and \citet{CvB12}.
These compare favorably to heating rates determined from {\em Helios}
and {\em Ulysses} measurements \citep{Cm09}, which are also
illustrated in Figure \ref{fig10}.
These heating rates are all roughly consistent with a power-law
scaling of $Q \propto r^{-4.5}$.

Figure \ref{fig10} also shows two calculations of the radial
dependence of $Q_{p}$ that assumed proton cyclotron resonance is
the sole source of heating.
For the fast-wind (polar coronal hole) model of \citet{CvB12}---which
specifies time-steady background quantities such as $V_{\rm A}$ and
$\beta$---we solved Equation (\ref{eq:Idef}) for $Q_p$.
This requires knowledge of both the wave power normalization
quantity $P_1$ and the total scaled heating rate
$I_{p} = I_{p \parallel}+I_{p \perp}$.
Assuming isotropic proton VDFs (${\cal R}=1$), the only other
parameter that causes $I_p$ to vary is $\beta$.
We extracted this dependence from the model
illustrated in Figure \ref{fig08} and fit it with
\begin{equation}
  \log I_{p} \, \approx \, -0.324 + 0.227 \log\beta +
  0.0168 (\log \beta)^{2}
  \,\,\,\,\,\, (\mbox{for} \,\, {\cal R}=1) .
\end{equation}
To obtain the red and blue curves in Figure \ref{fig10}, $P_{1}$
was estimated in two different ways.
First, we took as an upper limit the isotropic power spectrum
model of Equation (\ref{eq:PBnodamp}), with the quantities
$\langle \delta B_{\perp}^{2} \rangle$ and $k_0$ taken from the
wave transport model of \citet{CvB12}.
This model is labeled ``isotropic $P_{\rm B}$'' in Figure \ref{fig10},
and it produces several orders of magnitude {\em greater} proton
heating than is inferred to exist in the fast wind.
This result is consistent with the results of \citet{IV11}, who
found that one needs only $\sim 10^{-2}$ of the total available
wave power to be in the form of high-$k_{\parallel}$ cyclotron waves
in order to heat the protons adequately.

On the other hand, the blue curve labeled ``damped/mode-coupled
$P_{\rm B}$'' was computed using the full \citet{CvB12} model for
the Alfv\'{e}nic power spectrum in the corona and heliosphere.
This model contained an anisotropic cascade, with only a
small fraction of the total wave energy reaching the
high-$k_{\parallel}$ cyclotron resonance.
In Figure \ref{fig10}, we plot only the proton heating rate due to
cyclotron wave damping, but the original model also contained
proton heating due to Landau and transit-time resonances.
Because the high-$k_{\parallel}$ waves were strongly depleted in
this model (relative to an isotropic power spectrum), the
heating rate $Q_p$ is much smaller than is generally believed to
be needed to heat solar wind protons.
A complete understanding of proton energetics is likely to require
more than one source of heat.
In Section \ref{sec:radial} we assume that $Q_{p \parallel}$ and
$Q_{p \perp}$ are given by linear combinations of terms from
cyclotron resonance and an unspecified second source that, for
simplicity, we assume heats the protons isotropically.

\section{The Effect of Drifting Alpha Particles}
\label{sec:alpha}

When multiple ion species are present in a plasma containing cyclotron
resonant waves, it is possible for some ions to block others from
receiving the full extent of the heating they would have received
in isolation.
This effect has been studied extensively for the case of alpha
particles preventing the protons from being heated resonantly
\citep[e.g.,][]{Lw01,Xie4,Ga05,Mn13,Kp13}.
Also, \citet{Cr00,Cr01} found that even minor ions (with, e.g.,
$n_{i}/n_{p}$ as low as $\sim$10$^{-5}$) may be efficient at
absorbing high-$k_{\parallel}$ waves that propagate up from the
solar surface and become resonant high in the corona.
A main conclusion from these studies has been that some kind of
gradual replenishment of the wave spectrum---such as from a
turbulent cascade---is needed to explain how the protons may be
heated in this way.
Sometimes, however, the resonant damping may be so rapid that even an
efficient cascade may not be able to supply wave power to the protons.
The goal of this section is to estimate the degree to which proton
heating rates ($Q_{p \parallel}$, $Q_{p \perp}$) are suppressed by
the presence of alpha particles.

The models described in Section \ref{sec:heat} assumed a power-law
form for $P_{\rm B}(k_{\parallel})$.
Here, we aim to take into account the high-$k_{\parallel}$ damping
due to both protons and alphas in a more self-consistent way.
We follow \citet{CvB12} and assume the ``replenishment'' of the
Alfv\'{e}nic fluctuation spectrum comes from an isotropic cascade of
fast-mode waves that are locally mode-converted into Alfv\'{e}n waves.
This last step is assumed to be relatively instantaneous, which
enables us to combine the effects of cascade and damping into a
single transport equation for the Alfv\'{e}nic power spectrum.
This transport equation is written in terms of the full 3D power
spectrum $E_{\rm A}({\bf k})$ to retain continuity with the equations
of \citet{CvB12}.

In the past, turbulent cascades have been modeled as a combination
of wavenumber {\em advection} (i.e., first-order transport that goes
strictly from low to high $k$) and {\em diffusion} (i.e.,
second-order transport that spreads out the power in both directions,
but ends up ultimately with a turbulent power law).
Many aspects of the transport do not depend on the relative strengths
of the advection and diffusion terms, so for simplicity we assume
pure advection (see also Appendix C.2 of \citet{CvB12}).
Thus, the proposed transport equation, for an isotropic cascade in
$k$, is given by
\begin{equation}
  \frac{\partial E_{\rm A}}{\partial t} \, = \,
  -\frac{\mu}{k^2} \frac{\partial}{\partial k} \left(
  \frac{k^{3} E_{\rm A}}{\tau_c} \right) + 2\gamma E_{\rm A} \,\, .
  \label{eq:dEdt}
\end{equation}
The first term on the right-hand side describes the wavenumber
advection, where $\mu$ is an order-unity constant and $\tau_c$
is a $k$-dependent cascade timescale given by
\begin{equation}
  \tau_{c} \, = \, \frac{V_{\rm A}}{k v^2} \, = \,
  \frac{\rho_{0} V_{\rm A}}{k^{4} E_{\rm A}} \,\, .
  \label{eq:tauc}
\end{equation}
Equation (\ref{eq:tauc}) assumes that the timescale is constrained
by the weak Iroshnikov--Kraichnan type cascade experienced by
fast-mode waves prior to being coupled back to the Alfv\'{e}n mode.
The spectrum of velocity fluctuations $v(k)$ is related to the
energy spectrum as $\rho_{0} v^{2} = k^{3} E_{\rm A}$.

The second term on the right-hand side of Equation (\ref{eq:dEdt})
produces damping when $\gamma < 0$.
We assume that the resonant waves of interest are sufficiently
close to parallel propagation that $k \approx k_{\parallel}$.
The 1D transport equation is then solved by integrating from an
initial condition at an outer-scale wavenumber $k_0$.
This wavenumber is assumed to be far below the scales at which the
resonant damping occurs.
Thus, for a time-steady system ($\partial E_{\rm A} / \partial t = 0$),
the solution of Equation (\ref{eq:dEdt}) is
\begin{equation}
  E_{\rm A}(k) = E_{0} \left( \frac{k_0}{k} \right)^{7/2} +
  \frac{\rho_{0} V_{\rm A}}{\mu k^{7/2}} \int_{k_0}^{k} 
  \frac{dk}{k^{3/2}} \, \gamma(k)  \,\, ,
  \label{eq:Edampsol}
\end{equation}
where $E_0$ is defined as the known power level at $k_0$, and it
is related to the root-mean-squared fluctuation velocity via
$k_{0}^{3} E_{0} = \rho_{0} \langle \delta v_{\perp}^{2} \rangle$.
Using the equations from Section \ref{sec:heat:spec}, the 1D
reduced power spectrum is given by
$P_{\rm B} = 2\pi k_{\parallel}^{2} E_{\rm A}/3$.

Equation (\ref{eq:Edampsol}) exhibits an undamped inertial range at
low wavenumbers, and it steepens to an infinitely sharp asymptote near
the point at which the cascade timescale equals the inverse damping
rate (i.e., $|\gamma \tau_{c}| \approx 1$).
This transition from the inertial range to the dissipation range
appears to model the expected behavior of the system reasonably well,
despite the fact that the solution for $E_{\rm A}$ at even larger
wavenumbers is negative and unphysical.\footnote{%
A proper treatment of the {\em combined} effects of wavenumber
advection and diffusion would produce a more realistic exponential-like
decline in the dissipation range; see, e.g., Appendix C.5 of
\citet{CvB12}.}
For convenience, we express this solution dimensionlessly by dividing
by the inertial range solution,
\begin{equation}
  \widetilde{E} \, \equiv \,
  \frac{E_{\rm A}}{E_{0} (k_{0}/k)^{7/2}} \, = \,
  1 \, + \, C \int_{x_0}^{x} \frac{y \,\, dx}{x^{3/2}} 
  \left( \frac{\gamma}{\omega_r} \right)
\end{equation}
where as above we define $x = k_{\parallel} V_{\rm A}/\Omega_p$
and $y = \omega_{r}/\Omega_p$.
The key constant that sets the level of relative ``competition''
between cascade and damping is $C$, which is defined as
\begin{equation}
  C \, = \, \frac{8\pi V_{\rm A}^2}
  {\mu \langle \delta v_{\perp}^{2} \rangle}
  \left( \frac{\Omega_p}{k_{0} V_{\rm A}} \right)^{1/2} \,\, .
\end{equation}
Note that $C$ can also be written as the product of $\Omega_p$
and a representative cascade timescale that applies at $x \approx 1$.
In the solar wind, $C \gg 1$ because $k_0$ is always several 
orders of magnitude smaller than the resonant wavenumber
$\Omega_{p}/V_{\rm A}$.
Also, $V_{\rm A}^{2} \gg \langle \delta v_{\perp}^{2} \rangle$
in the corona, but these velocities are of the same order of magnitude
at larger heliocentric distances.
Using the fast solar wind model of \citet{CvB12}, we assumed $\mu = 2$
and we computed $C$ as a function of distance.
In the low corona, $C \gtrsim 10^{6}$, and it declines rapidly
to $\sim 10^{5}$ at $r = 4 \, R_{\odot}$,
to $\sim 10^{4}$ at $r = 20 \, R_{\odot}$, and
to $\sim 10^{3}$ at $r = 5$~AU.

\begin{figure}
\epsscale{1.15}
\plotone{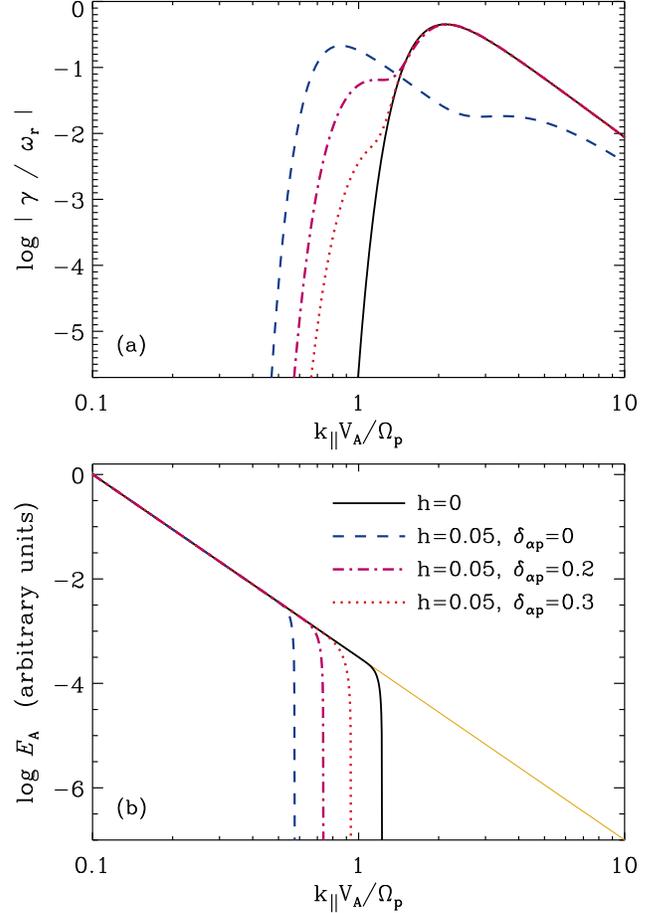}
\caption{(a) Absolute values of dimensionless damping rates
$|\gamma / \omega_{r}|$ as a function of parallel wavenumber.
(b) Damped wave spectra consistent with damping rates shown
in panel (a), compared with the undamped spectrum (solid gold curve).
In both panels, the result for a pure proton-electron plasma
(black solid curve) is compared with that for 5\% helium abundance
and drift speeds $\delta_{\alpha p} = 0$ (blue dashed curve),
0.2 (magenta dot-dashed curve), and 0.3 (red dotted curve).
\label{fig11}}
\end{figure}

Figure \ref{fig11} shows how the damping rates and resulting spectra
change when the helium abundance $h$ and relative drift speed
$\delta_{\alpha p}$ are varied.
The damping rate $\gamma$ is computed by solving
Equation (\ref{eq:gow}), assuming bi-Maxwellian VDFs for both protons
and alpha particles, and Equation (\ref{eq:cdispA}) was used for the
real part of the dispersion relation.
In Figure \ref{fig11}, the parameters held fixed are 
$\beta = 0.01$, $C = 10^{4}$, $T_{\alpha}/T_{p} = 4$, and
${\cal R}=1$ for both protons and electrons.
Because of the lower charge-to-mass ratio of the alphas, they have
the opportunity to undergo cyclotron resonance at lower values of
$k_{\parallel}$ than do the protons.

For $\beta \ll 1$, the power spectrum computed for no alphas
($h=0$) is indistinguishable from that computed for $h=0.05$
and $\delta_{\alpha p} \approx 1$.
For large drifts, the alphas become ``Doppler shifted''
well out of resonance and the dispersion relation is effectively
that of a pure proton--electron plasma.
However, at higher values of $\beta$, \citet{Vs13} found that the
presence of drifting alphas may destabilize the protons
near $\delta_{\alpha p} \approx 1$, leading to $\gamma > 0$ and
the possibility of wave growth.
This effect was also found in the models discussed here, but only
for a narrow range of drift parameters $\delta_{\alpha p}$ extremely
close to 1.

For the case of $h=0.05$ and several fixed choices of $C$, damped
power spectra were computed for a large grid of values of $\beta$
and $\delta_{\alpha p}$.
Each spectrum was processed through Equation (\ref{eq:biheat}) to
obtain heating rates $Q_{p \parallel}$ and $Q_{p \perp}$.
We found that a convenient dimensionless way to measure the
ability of alpha particles to suppress the proton heating
(as derived in Section \ref{sec:heat}) is the ratio
\begin{equation}
  {\cal D} (\beta, \delta_{\alpha p}) \, = \,
  \frac{Q_{p \perp}(\beta, \delta_{\alpha p})}{Q_{p \perp}(\beta, 1)}
  \,\, .
  \label{eq:Ddef}
\end{equation}
Since ${\cal D} \leq 1$, this ratio describes how a given level of
alpha--proton drift gives rise to an additional amount of damping
relative to what would occur when the alphas are fully out of resonance.
The behavior for $Q_{p \parallel}$ is nearly identical to the behavior
for $Q_{p \perp}$, so for simplicity we use Equation (\ref{eq:Ddef})
for both.

\begin{figure}
\epsscale{1.16}
\plotone{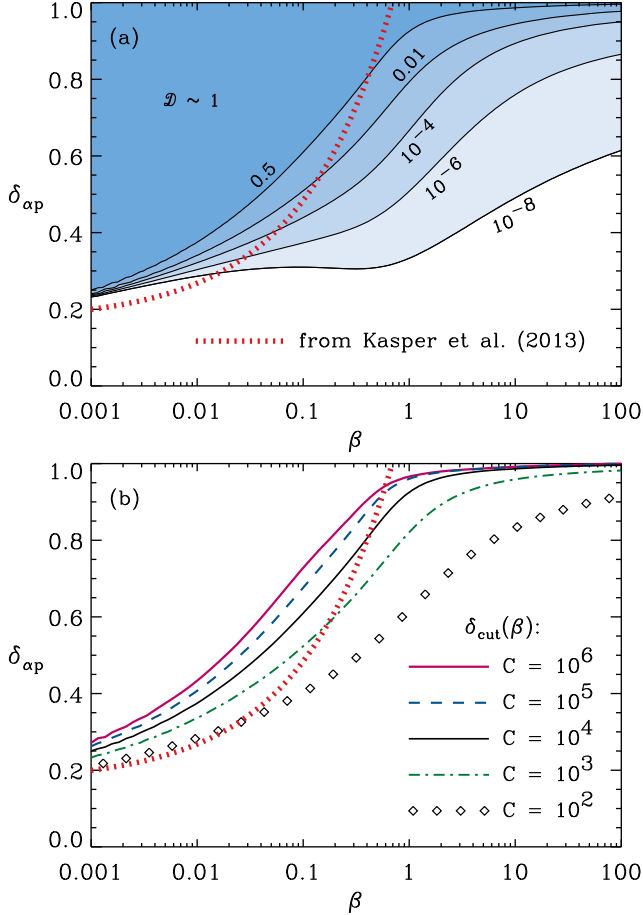}
\caption{(a) Contours of constant values of ${\cal D}$ as a function
of proton $\beta$ and alpha--proton drift speed $\delta_{\alpha p}$,
computed for $C = 10^4$.
(b) $\delta_{\rm cut}$ plotted versus $\beta$ for models computed
with a range of cascade--damping constants $C$ (see labels).
In both panels, the red dotted curve corresponds
to Equation (\ref{eq:kasper}).
\label{fig12}}
\end{figure}

Figure \ref{fig12}(a) shows the full dependence of ${\cal D}$ on
on both $\beta$ and $\delta_{\alpha p}$ for 
$C = 10^{4}$ (representative of the inner heliosphere)
and fixed values of ${\cal R} = 1$ and $T_{\alpha}/T_{p} = 4$.
There is a plateau of ${\cal D} \approx 1$ at low values of $\beta$
and large values of $\delta_{\alpha p}$.
However, ${\cal D}$ drops precipitously as one moves from the
upper-left to the lower-right part of the panel.
As suspected from earlier studies \citep[e.g.,][]{Ga05,Bu11},
alpha particles at low drift speeds capture most of the wave
energy themselves and prevent the protons from being heated
(i.e., ${\cal D} \approx 0$).
Figure \ref{fig12} also shows the dividing line given by
\citet{Kp13}, which was parameterized as
\begin{equation}
  \delta_{\rm cut, K} \, = \,
  \min \left( 0.168 + \beta^{1/2} , 1 \right)
  \label{eq:kasper}
\end{equation}
and was proposed to account for the $\beta$-dependence of the
resonant cutoff between regions of strong and weak alpha particle
suppression of proton heating.
This curve is quite similar in shape and position to the contours
that describe the dropoff of ${\cal D}$.

In order to make efficient use of the two-dimensional distributions
${\cal D} (\beta, \delta_{\alpha p})$ in spatially extended
solar wind models, we parameterized these functions as follows.
For a given grid of ratios like that plotted in Figure \ref{fig12}(a),
we determined the locus of points that corresponds to the
${\cal D} = 0.5$ contour.
This describes a function $\delta_{\rm cut} (\beta)$ that roughly
divides the grid into two regions of strong and weak alpha
suppression.
Figure \ref{fig12}(b) shows $\delta_{\rm cut}$ versus $\beta$ for
several grids that were computed with different values of $C$.
In addition, Table \ref{table01} provides $\delta_{\rm cut}$
for a coarse grid of $\beta$ values.
Given the numerical tabulation of $\delta_{\rm cut} (\beta)$, we
then represent the full dependence of ${\cal D}$ with
\begin{equation}
  {\cal D} (\beta, \delta_{\alpha p}) \, \approx \,
  \frac{1}{2} \left[ 1 + \mbox{erf} \left(
  \frac{\delta_{\alpha p} -
  \delta_{\rm cut}(\beta)}{\sigma_{\rm cut}} \right) \right]
  \label{eq:calDfit}
\end{equation}
where $\sigma_{\rm cut} = 0.04$ reproduces the numerical results
quite well.
For simplicity, in the solar wind simulations described below we
use only this model for a single intermediate value of $C = 10^4$.

\begin{deluxetable}{cccc}
\tablecaption{Cutoff Drift Speeds for Suppression of Proton Heating
\label{table01}}
\tablewidth{0pt}
\tablehead{\colhead{$\beta$} &
\colhead{$\delta_{\rm cut}$ ($C = 10^5$)} &
\colhead{$\delta_{\rm cut}$ ($C = 10^4$)} &
\colhead{$\delta_{\rm cut}$ ($C = 10^3$)}}
\startdata
   1E--3  &   0.2624  &   0.2501  &   0.2327 \\
   2E--3  &   0.2961  &   0.2786  &   0.2572 \\
   5E--3  &   0.3522  &   0.3283  &   0.2978 \\
   1E--2  &   0.4071  &   0.3748  &   0.3365 \\
   2E--2  &   0.4735  &   0.4327  &   0.3818 \\
   5E--2  &   0.5813  &   0.5253  &   0.4556 \\
   1E--1  &   0.6763  &   0.6098  &   0.5234 \\
   2E--1  &   0.7732  &   0.7025  &   0.6011 \\
   5E--1  &   0.9052  &   0.8394  &   0.7235 \\
   1E+0   &   0.9601  &   0.9248  &   0.8206 \\
   2E+0   &   0.9776  &   0.9630  &   0.8928 \\
   5E+0   &   0.9865  &   0.9806  &   0.9412 \\
   1E+1   &   0.9911  &   0.9864  &   0.9588 \\
   2E+1   &   0.9942  &   0.9906  &   0.9692 \\
   5E+1   &   0.9972  &   0.9942  &   0.9778 \\
   1E+2   &   0.9990  &   0.9959  &   0.9820 \\
\vspace*{0.10in}
\enddata
\end{deluxetable}

\section{Radial Evolution of the Proton Velocity Distribution}
\label{sec:radial}

Earlier sections described how solar wind protons are energized
by cyclotron resonant interactions in regions where the background
plasma parameters are assumed to be fixed and homogeneous.
Here we develop a larger-scale global model of how these effects may
manifest themselves along flux tubes that extend from the corona
($r \approx 2 \, R_{\odot}$) to interplanetary space.
Section \ref{sec:radial:eqns} lays out the bi-Maxwellian moment
equations adopted for this model.
We acknowledge that real heliospheric VDFs are never exactly
bi-Maxwellian, but we wish to explore the extent to which the
moment equations faithfully model the kinetic processes that
dominate proton energetics in the solar wind.
Results are presented first for a flux tube representative of
high-latitude fast wind streams (Section \ref{sec:radial:result1}),
then for a broad Monte Carlo ensemble of heliospheric
parameters (Section \ref{sec:radial:result2}).

\subsection{Conservation Equations}
\label{sec:radial:eqns}

The radial evolution of proton VDFs is modeled here by time-steady
1D conservation equations for the bi-Maxwellian temperature
parameters $T_{p \parallel}$ and $T_{p \perp}$ and a simplified
equation for the alpha--proton drift parameter
$\delta_{\alpha p}$.
The radial dependences of proton number density $n_p$ and outflow
speed $u_p$, as well as electron temperature $T_e$, are determined
separately (see below).
The temperature equations were simplified from the 16-moment
model of \citet{Li99} and are given by
\begin{displaymath}
  \frac{\partial T_{p \parallel}}{\partial r} \, = \,
  -T_{p \parallel} \left( \frac{2 \cos^{2} \Phi}{L_u} +
  \frac{\sin^{2} \Phi}{L_b} \right)
\end{displaymath}
\begin{equation}
  + \, \frac{4 \nu_{pp}}{5 u_p}
  \left( T_{p \perp} - T_{p \parallel} \right) +
  \frac{2 \nu_{pe}}{u_p} \left( T_{e} - T_{p \parallel} \right) +
  \frac{2 Q_{p \parallel}}{n_{p} u_{p} k_{\rm B}}
  \label{eq:dTparadr}
\end{equation}
\begin{displaymath}
  \frac{\partial T_{p \perp}}{\partial r} \, = \,
  -T_{p \perp} \left( \frac{\sin^{2} \Phi}{L_u} +
  \frac{2 - \sin^{2} \Phi}{2 L_b} \right)
\end{displaymath}
\begin{equation}
  + \, \frac{2 \nu_{pp}}{5 u_p}
  \left( T_{p \parallel} - T_{p \perp} \right) +
  \frac{2 \nu_{pe}}{u_p} \left( T_{e} - T_{p \perp} \right) +
  \frac{Q_{p \perp}}{n_{p} u_{p} k_{\rm B}}
  \label{eq:dTperpdr}
\end{equation}
\citep[see also][]{IH83,CFK99,Me12}.
From left to right, terms on the right-hand sides of
Equations (\ref{eq:dTparadr})--(\ref{eq:dTperpdr}) describe
double-adiabatic expansion, collisional isotropization,
electron--proton collisional equilibration, and net heating.
These equations do not include proton heat conduction, which is often
found to be of negligible importance in solar wind thermodynamics
\citep{SL95,Cm09,Hg13}.
Two key scale lengths used above are defined as
\begin{equation}
  \frac{1}{L_u} = \frac{1}{u_p} \frac{\partial u_p}{\partial r}
  \,\,\, , \,\,\,\,\,\,\,\,
  \frac{1}{L_b} = -\frac{1}{B_r} \frac{\partial B_r}{\partial r}
  \,\, .
\end{equation}
For models at low heliographic latitudes, we include the effect
of the Parker spiral by defining
\begin{equation}
  \tan \Phi \, = \, \frac{B_{\phi}}{B_r} \, = \,
  - \frac{\Omega_{s} r \sin\vartheta}{u_{1{\rm AU}}}
  \label{eq:spiral}
\end{equation}
where $\vartheta$ is the colatitude of the wind streamline
(usually $\pi/2$ for the ecliptic) and the solar
rotation rate is $\Omega_{s} = 2.7 \times 10^{-6}$ rad s$^{-1}$.
Equation (\ref{eq:spiral}) uses $u_{1{\rm AU}}$, the wind speed at 1~AU,
instead of the radially varying wind speed, to better approximate the
end-result of solving the full set of MHD angular momentum equations
\citep{WD67}.

The proton--proton Coulomb collision rate was given by \citet{Li99} as
\begin{equation}
  \nu_{pp} \, = \, \frac{4}{3} \sqrt{\frac{\pi}{m_p}}
  \frac{n_{p} e^{4} \ln\Lambda}{(k_{\rm B} T_{p})^{3/2}} \,\, ,
\end{equation}
with the one-fluid proton temperature defined as
$T_{p} = (T_{p \parallel} + 2 T_{p \perp})/3$ and the Coulomb
logarithm given approximately by
\begin{equation}
  \ln \Lambda \, = \, 23.2 + \frac{3}{2} \ln \left(
  \frac{T_p}{10^{6} \, \mbox{K}} \right) - \frac{1}{2} \ln \left(
  \frac{n_p}{10^{6} \, \mbox{cm}^{-3}} \right)
\end{equation}
\citep[see, e.g.,][]{CvB07}.
Similarly, the electron--proton collision rate is
\begin{equation}
  \nu_{pe} \, = \, \frac{4}{3} \frac{\sqrt{2\pi m_{e}}}{m_p}
  \frac{n_{p} e^{4} \ln\Lambda}{(k_{\rm B} T_{e})^{3/2}} \,\, .
\end{equation}
The above expressions do not take account of temperature anisotropy
effects on the collision rates \citep[see, e.g.,][]{BS82,HT09}.
A useful quantity for studying solar wind parcels at 1~AU is
the dimensionless proton {\em collisional age} $A_p$.
This quantity is often defined as the product of the proton
self-collision rate and an estimate of the solar wind transit-time
from the Sun to a given radius.
Defining the latter as $t_{\rm wind} \approx r / u_{p}$, we define
$A_{p}  = \nu_{pp} t_{\rm wind}$,
and we evaluate these quantities all at 1~AU.
Protons in wind streams with $A_{p} \gg 1$ have experienced many
collisions and should be well isotropized.

The proton heating rates $Q_{p \parallel}$ and $Q_{p \perp}$ contain
the distilled results of the models developed in Sections
\ref{sec:disp}--\ref{sec:alpha}.
Figure \ref{fig10} shows that various models of turbulent transport
tend to produce power-law radial dependences $Q \propto r^{-n}$.
The models presented below use this simple form as a starting point.
Although we acknowledge that the actual heating rates must be
functions of the local turbulence amplitudes, correlation lengths,
and cascade rates, we also want to focus here on the relative
{\em partitioning} of heat between various modes of wave-particle 
interaction as described above.
Thus, we believe that treating the total available heat as a sum
of power-law components may not be too unrealistic.

Two power-law heating components are utilized: one due to ion cyclotron
resonance, and one that is assumed to heat the protons isotropically.
This latter source is proposed mainly because it is a simple
``null hypothesis,'' not because of the existence of any single
mechanism that produces isotropic heating.
Still, if there exist several other heating processes in the solar
wind besides ion cyclotron resonance---with multiple steps of energy
conversion prior to the final step of particle heating---it may not
be unrealistic to assume their summed effect provides comparable
amounts of heat to $T_{p \parallel}$ and $T_{p \perp}$.
The radial dependences of the two total rates are given as
\begin{equation}
  Q_{\rm cyc} \, = \, Q_{1} \left(
  \frac{R_{\odot}}{r} \right)^{\psi_1}
  \,\,\, , \,\,\,\,\,\,\,
  Q_{\rm iso} \, = \, Q_{2} \left(
  \frac{R_{\odot}}{r} \right)^{\psi_2}
  \label{eq:Qpower}
\end{equation}
where the normalizing constants $Q_1$ and $Q_2$ and the exponents
$\psi_1$ and $\psi_2$ are free parameters.
Once these are specified, the parallel and perpendicular heating
rates are
\begin{equation}
  Q_{p \parallel} \, = \, \frac{Q_{\rm cyc} {\cal D} I_{p \parallel}}
  {| I_{p \parallel} + I_{p \perp} |} + \frac{Q_{\rm iso}}{3}
  \label{eq:Qradpara}
\end{equation}
\begin{equation}
  Q_{p \perp} \, = \, \frac{Q_{\rm cyc} {\cal D} I_{p \perp}}
  {| I_{p \parallel} + I_{p \perp} |} + \frac{2 Q_{\rm iso}}{3}
  \label{eq:Qradperp}
\end{equation}
where $I_{p \parallel}$ and $I_{p \perp}$ are specified by
Equation (\ref{eq:biI}) and ${\cal D}$ is given by
Equation (\ref{eq:calDfit}).
The factors of $1/3$ and $2/3$ in the $Q_{\rm iso}$ terms are
there to ensure that $T_{p \parallel}$ and $T_{p \perp}$ would
receive equal rates of increase if there were no other heating or
cooling terms in Equations (\ref{eq:dTparadr})--(\ref{eq:dTperpdr}).

To avoid a proliferation of free parameters, the exponent
$\psi_{2}$ is set to a constant value of 4.5.
This is consistent with the implicit assumption that $Q_{\rm iso}$
is dominated by the dissipation of a perpendicular KAW cascade
(see Figure \ref{fig10}).
The other three parameters ($\psi_1$, $Q_1$, $Q_2$) are varied
freely in the models.
In practice, however, we select $Q_1$ and the value of 
$Q_{\rm cyc}$ at 1~AU, and then solve for the value of $\psi_1$ that
connects them with a power law.
The kinetic effects of cyclotron wave damping and instability are
included in Equations (\ref{eq:Qradpara})--(\ref{eq:Qradperp})
because both $I_{p \parallel}$ and $I_{p \perp}$ depend on the
local values of ${\cal R}$ and $\beta$ as illustrated in
Figure \ref{fig08}.
The effect of the firehose instability is not included explicitly in
$Q_{p \parallel}$ and $Q_{p \perp}$, but in
Section \ref{sec:radial:result1} we discuss an approximate method
of computing its net impact on the proton VDF.

The dimensionless alpha--proton drift parameter $\delta_{\alpha p}$
should be determined from a complete solution of the coupled momentum
transport equations for $u_{\alpha}$ and $u_{p}$.
For now, however, we make a first step in this direction by
solving a simpler radial evolution equation.
This equation contains only the collisional friction that is expected
to produce a steady decrease in $\delta_{\alpha p}$; i.e.,
it starts at a specified initial condition in the corona, and
it is driven toward zero in the limit of a strongly collisional
history.
The differential momentum equation of \citet{Hz87} was adapted into
the following form,
\begin{equation}
  \frac{\partial}{\partial r} \left( \ln \delta_{\alpha p} \right)
  \, = \, -\frac{\nu_{\alpha p}}{u_p} \left[
  \frac{\mbox{erf}(\zeta) - 2\zeta e^{-\zeta^2}/\sqrt{\pi}}
  {\zeta^3} \right]
  \label{eq:ddeldr}
\end{equation}
where an effective drift Mach number is defined as
\begin{equation}
  \zeta \, = \, \delta_{\alpha p} V_{\rm A} \left(
  \frac{2k_{\rm B} T_{\alpha}}{m_{\alpha}} + 
  \frac{2k_{\rm B} T_p}{m_p} \right)^{-1/2}
\end{equation}
and the frictional collision rate is
\begin{equation}
  \nu_{\alpha p} \, = \, 16 \sqrt{2} \, \nu_{pp} \,
  \left( \frac{T_{\alpha}}{T_p} + 4 \right)^{-3/2}  \,\, .
\end{equation}
When $\zeta \ll 1$, the square-bracket term in
Equation (\ref{eq:ddeldr}) approaches a constant value of $\sim$0.752.
When $\zeta \gg 1$ (i.e., the runaway regime), the term in square
brackets declines as $1/\zeta^3$.
As above, we assume $T_{\alpha}/T_{p} = 4$ to evaluate the above terms.
Conveniently, in this limit, $\nu_{\alpha p} = \nu_{pp}$.

Equations (\ref{eq:dTparadr}), (\ref{eq:dTperpdr}), and
(\ref{eq:ddeldr}) are integrated from low to high $r$ with
straightforward first-order Euler steps.
We chose a lower boundary of $r = 2 \, R_{\odot}$ at which
the proton VDF is assumed to be isotropic and in
collisional equilibrium with the electron VDF.
Thus, the lower boundary condition is that
$T_{p \parallel} = T_{p \perp} = T_{e}$, where the latter is
given by an empirical parameterization that agrees reasonably well
with coronal and in~situ measurements,
\begin{equation}
  T_{e} \, = \, \frac{10^{6} \,\, \mbox{K}}
  {0.3 (r/R_{\odot})^{0.6} \, + \, (r/R_{\odot})^{-1}} \,\, .
\end{equation}
This expression has a maximum value of $\sim$1.1 MK at
$r \approx 2.9 \, R_{\odot}$, and it decreases slowly to about
0.13 MK at 1~AU.
The lower boundary condition on the alpha--proton drift parameter
is fixed at $\delta_{\alpha p} = 1$, which assumes strong differential
acceleration of heavy ions in the corona, as inferred from
coronagraph spectroscopy \citep{Ko06}.

At heliocentric distances above 2 $R_{\odot}$, the radial magnetic
field strength is reasonably approximated by a spherically expanding
flux tube, with $B_{r} \propto r^{-2}$ and thus $L_{b} = r/2$.
The azimuthal field strength $B_{\phi}$ was computed from
Equation (\ref{eq:spiral}) and the vector magnitude
$B_{0} = (B_{r}^{2} + B_{\phi}^{2})^{1/2}$ was used in the
definitions of $V_{\rm A}$ and $\Omega_p$.
The solar wind is still undergoing significant acceleration in
the regions around $r = 2 \, R_{\odot}$, so it was not reasonable
to assume a constant value of $u_{p}$.
We found that the low-latitude solar wind models of \citet{CvB13}
all exhibited a roughly similar {\em relative} acceleration
profile above 2 $R_{\odot}$, which we fit by a simple function
\begin{equation}
  \frac{u_{p} (r)}{u_{1{\rm AU}}} \, = \,
  0.062 + 0.938 \left( 1 - \frac{R_{\odot}}{r} \right)^{5.1}
  \,\, . \label{eq:uofr}
\end{equation}
An actual model of $u_{p}(r)$ is then obtained by choosing a
normalizing value of $u_{1{\rm AU}}$, which we do either by an
explicit choice or by sampling from a random probability
distribution.
The derivative of Equation (\ref{eq:uofr}) is used to find the
radial dependence of $L_u$.
Lastly, mass flux conservation is used to obtain the radial
dependence of density.
We set its absolute value by choosing the sphere-averaged mass
loss rate $\dot{M}$, and at 1~AU the density is computed as
\begin{equation}
  \rho_{1{\rm AU}} \, = \,
  \frac{\dot{M}}{(4 \pi u_{p} r^{2})_{1{\rm AU}}} \,\, .
  \label{eq:rho1au}
\end{equation}
At other heights, $\rho(r)$ obtained from the assumption that
$\rho u / B_{r}$ remains constant.
For the models described below, we assumed $h=0.05$ and we converted
between mass density and proton number density with
$\rho = m_{p} n_{p} (1 + 4h)$.

\subsection{Results: Fast Solar Wind}
\label{sec:radial:result1}

To begin exploring how the conservation equations behave
when the anisotropic heating rates are varied, we set the
background solar wind conditions to those appropriate for a
high-speed stream connected to a polar coronal hole
\citep[see, e.g., Section 2.1 of][]{CvB12}.
Specifically, we set
$u_{1{\rm AU}} = 750$ km s$^{-1}$,
$B_{r,1{\rm AU}} = 3$ nT, and
$\dot{M} = 2 \times 10^{-14}$ $M_{\odot}$ yr$^{-1}$.
All other parameters were determined as described in
Section \ref{sec:radial:eqns}.

\begin{figure}
\epsscale{1.15}
\plotone{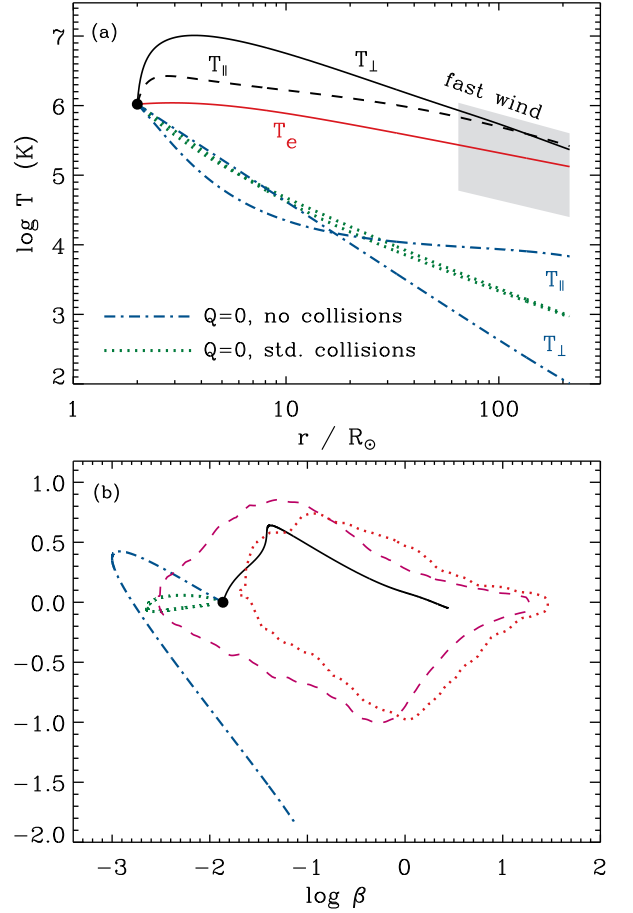}
\caption{(a) Radial dependence of $T_{p \parallel}$ and $T_{p \perp}$
for fast-wind models, shown together with (b) associated curves in
($\beta$,${\cal R}$) space.
Initial conditions shown as black circles.
Proton temperatures computed with strong heating (black curves)
are contrasted with models computed for $Q_{p}=0$ with no
collisions (blue dot-dashed curves) and for $Q_{p}=0$
with standard collision rates (green dotted curves).
Also shown is the empirical $T_e$ (red solid curve).
A gray region in (a) encloses the full range of $T_p$ variation
seen by {\em Helios}, and red/magenta regions in (b) show
the range of {\em Wind} data at 1~AU.
\label{fig13}}
\end{figure}

Figure \ref{fig13}(a) shows the radial dependence of $T_{p \parallel}$
and $T_{p \perp}$ for several cases, and Figure \ref{fig13}(b)
shows corresponding trajectories in the beta--anisotropy plane.
As has been known for several decades \citep[e.g.,][]{HS68},
a complete lack of proton heating gives rise to proton temperatures
well below the measured values at 1~AU.
When the full Coulomb self-collision terms are used in
Equations (\ref{eq:dTparadr})--(\ref{eq:dTperpdr}), the low
temperatures produced in the $Q_{p}=0$ model give large values
of $\nu_{pp}$.
Thus, the system is rapidly driven toward isotropy, and the radial
decrease in temperature is close to adiabatic
($T_{p} \propto r^{-4/3}$).
In this case, Figure \ref{fig13}(b) shows that the $\beta$ ratio
does not vary substantially on its journey from the corona to
interplanetary space, and in fact at 1~AU it loops back to nearly
its initial value.

Figure \ref{fig13} also shows the result of turning off all
Coulomb collisions for the model with $Q_{p}=0$.
With neither heating nor collisional isotropization, the proton
VDF obeys strict magnetic moment conservation and is beamed to
a state of ${\cal R} \ll 1$ at 1~AU \citep[e.g.,][]{Ho71,LA72,Dn94}.
An even more idealized case of a constant wind speed, a radial
field, and no Parker spiral would produce
$T_{p \parallel} = \mbox{constant}$ and $T_{p \perp} \propto r^{-2}$.
This is close to what is seen in the blue curves of Figure \ref{fig13}(a)
above $r \approx 10 \, R_{\odot}$.

Lastly, Figure \ref{fig13} shows the result of applying realistic
proton heating parameters that were chosen to reproduce the
median {\em Helios} fast-wind measurements reported by \citet{Ma82}.
These conditions were notable because the temperatures were near
the high end of the full range of observed conditions, and
they exhibited ${\cal R} > 1$ at 0.3~AU and
${\cal R} < 1$ at 1~AU.
The three parameters that reproduce this state were found to be
$Q_{1} = 1.596 \times 10^{-6}$ erg s$^{-1}$ cm$^{-3}$,
$\psi_{1} = 4.150$, and
$Q_{2} = 1.660 \times 10^{-6}$ erg s$^{-1}$ cm$^{-3}$.
To generate the observed conditions, the model needed to have
$Q_{\rm cyc}/Q_{\rm iso} \approx 1$ in the corona, and then this
ratio needed to increase to about 6 at 1~AU.
Figure \ref{fig13}(b) shows that this model's coronal evolution
follows the left edge of one of the {\em Wind} beta--anisotropy
regions \citep{Ma12}.
At larger distances, the strong heating is necessary to drive the
model to larger values of $\beta$ as seen in the data.
For this model, $\delta_{\alpha p}$ decreased only negligibly from
its initial value of 1 to a value of 0.973 at 1~AU.
Below, we find that the more dominant slow/dense solar wind
undergoes a much stronger heliospheric decline in $\delta_{\alpha p}$.

The strongly heated fast-wind model shown in Figure \ref{fig13}
exhibits a total heating rate at 1~AU of
$Q_{p} = 3.87 \times 10^{-16}$ erg s$^{-1}$ cm$^{-3}$, which falls
comfortably within the uncertainty limits of the rates computed
from in~situ data by \citet{Vb07} and \citet{Cm09}.
The agreement with coronal observations is not so good;
the model gives a maximum value of $T_{p \perp} \approx 10$ MK
in the extended corona, whereas the Ultraviolet Coronagraph
Spectrometer (UVCS) instrument on the {\em Solar and Heliospheric
Observatory} ({\em{SOHO}}) allowed for no more than
$T_{p \perp} \approx 3$ MK \citep{Ko06}.
Note, however, that our use of a lower boundary condition at
$r = 2 \, R_{\odot}$ may be the cause of unrealistic conditions
in the first few solar radii above the surface.
We experimented with using a smaller initial radius, but a proper
model with $r < 2 \, R_{\odot}$ would necessitate the use of
departures from the power-law heating rates given in
Equation (\ref{eq:Qpower}).

Although none of the three models displayed in Figure \ref{fig13}
ventured into the firehose-unstable region in the lower-right part
of the ($\beta$,${\cal R}$) plane, other models are likely to do so.
Thus, for the results shown below we augment the solution algorithm
for $T_{p \perp}(r)$ in a way that accounts for this instability.
\citet{Me12} and \citet{Sg13} showed that if the system is driven
to a point below the curve of marginal firehose stability, the
net effect of the instability will be to heat up $T_{p \perp}$ and
leave $T_{p \parallel}$ relatively unchanged.
Thus, for $\beta > 2$ we impose a lower limit on the perpendicular
temperature that demands
\begin{equation}
  T_{p \perp} \, \geq \, \left( 1 - \frac{2}{\beta} \right)
  T_{p \parallel}
  \label{eq:fire}
\end{equation}
(see Section \ref{sec:disp:bimax} and Figure \ref{fig02}).
If the locally computed value of $T_{p \perp}$ ever violates the
above condition, we replace it with the lower-limit value that
depends on $\beta$ and $T_{p \parallel}$.
This is similar to the techniques employed by \citet{Sh06} and
\citet{Ch11}.
Equation (\ref{eq:fire}) is consistent with the classical
nonresonant firehose instability, but similar expressions taken
from calculations of the resonant parallel or oblique firehose
instability \citep{Ga98,Ro11,Ma12,Mi14} could also be used.

\begin{figure}
\epsscale{1.16}
\plotone{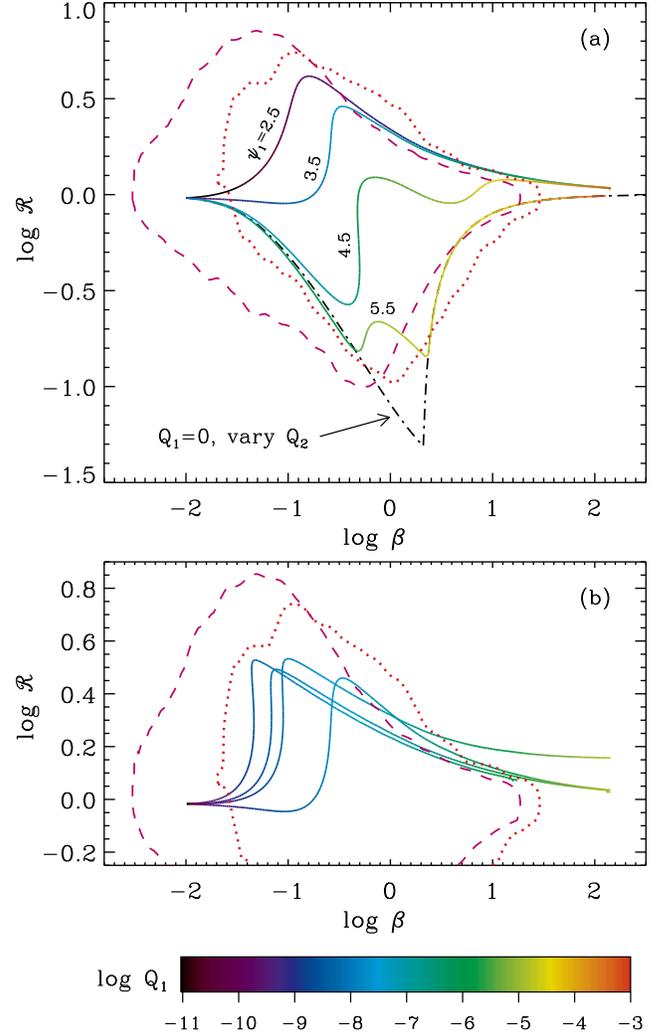}
\caption{Fast-wind results at 1~AU in the beta--anisotropy plane.
(a) Multicolor curves have variable $Q_1$ and fixed values of
$\psi_1$ (see labels and color bar) and also assume $Q_{2}=0$.
Black dot-dashed curve has variable $Q_2$ and assumes $Q_{1}=0$.
All curves in (a) use Equation (\ref{eq:yhotx}) for the dispersion
relation.
(b) Multicolor curves have variable $Q_1$,
$\psi_{1} = 3.5$, and $Q_{2}=0$.
From left to right (in the nearly vertical parts of the curves),
the corresponding dispersion relations are Equations (\ref{eq:wad}),
(\ref{eq:cdispP}), (\ref{eq:philfit}), and (\ref{eq:yhotx}).
\label{fig14}}
\end{figure}

In order to illustrate how the solutions depend on the heating
parameters $Q_1$, $\psi_1$, and $Q_2$, we created
several grids of models in which each of the parameters is varied
independently of the others.
Figure \ref{fig14} shows a selection of these model results
in the ($\beta$,${\cal R}$) plane.
It should be made clear that these curves do {\em not} show
radial variations (like in Figure \ref{fig13}) but instead give the
conditions at 1~AU that result from smoothly varying the heating
parameters.
Different points along each curve correspond to models with
different values of either $Q_1$ or $Q_2$, and an increase in
one of the heating rates tends to push the solutions from
left to right in the diagram.
We point out several properties of these models:
\begin{enumerate}
\item
The multicolor curves in Figure \ref{fig14}(a) show the result
of varying $Q_1$ and fixing $Q_{2} = 0$.
In other words, they show that it is possible to populate nearly
the entire observed region of the beta--anisotropy plane at 1~AU
by including {\em only} proton cyclotron resonant heating and
a sufficiently broad range of $Q_1$ and $\psi_1$ values.
Solar wind flux tubes with shallower radial power-law indices
(i.e., smaller values of $\psi_1$) give rise to strong
perpendicular anisotropies with ${\cal R} > 1$, and flux tubes
with steeper power-law indices produce ${\cal R} \lesssim 1$
at 1~AU.
\item
The black dot-dashed curve in Figure \ref{fig14}(a) shows the result
of varying $Q_2$ and fixing $Q_{1} = 0$.
It shows that purely isotropic proton heating does not provide
sufficient thermal energy to $T_{p \perp}$ to populate most of the
interior of the observed region of parameter space.
In fact, for relatively low heating rates, the isotropic model
produces roughly the same values of $\beta$ and ${\cal R}$ at 1~AU
as does the cyclotron model with the steepest values of $\psi_1$.
The combination of these two cases appears to reproduce the
lower-left edge of the populated parameter region quite well.
A hypothetical model with substantially {\em more} parallel than
perpendicular heating would extend even further to the lower-left.
Thus, the absence of such cases in the observations seems to
tell us that the solar wind never exhibits
$Q_{p \parallel} \gg Q_{p \perp}$.
\item
Figure \ref{fig14}(b) shows the result of varying the dispersion
relation, which affects the dependence of $I_{p \parallel}$ and
$I_{p \perp}$ on $\beta$ and ${\cal R}$.
Each of these models was computed with $\psi_{1} = 3.5$ and
$Q_{2} = 0$, and the corresponding dispersion relations are
referenced in the caption.
At the largest values of $Q_1$, the models approach their
respective marginal stability curves as also shown in
Figure \ref{fig09}.
In the remainder of this paper, we choose to use the simple ``hot''
dispersion relation of Equation (\ref{eq:yhotx}), whose marginal
stability curve was highlighted in Figure \ref{fig09}(c) with a
thick solid black curve.
Although this model is more simplistic than most others discussed
in Section \ref{sec:disp}, it appears to best reproduce the
shape of the upper-right boundary of observed parameter space.
As illustrated in Figure \ref{fig04}(a),  this model may also be
consistent with the perpendicularly ``stretched'' proton VDFs
that result from Fermi-like acceleration seen in the numerical
diffusion models.
\end{enumerate}

\subsection{Results: Monte Carlo Solar Wind Variations}
\label{sec:radial:result2}

Data from the {\em Wind} spacecraft were assembled over several years
to define the occupied regions of ($\beta$,${\cal R}$) parameter space
outlined in Figures \ref{fig02}, \ref{fig04}, \ref{fig09}, \ref{fig13},
and \ref{fig14}.
The measured solar wind conditions included both fast and slow streams.
In this section, we attempt to simulate a realistic statistical
ensemble of solar wind plasma states and predict its distribution of
anisotropic proton thermal properties.
The resulting Monte Carlo model involves randomly selecting three
background plasma parameters
($u_{1{\rm AU}}$, $B_{r,1{\rm AU}}$, and $\dot{M}$)
and three proton heating parameters
($Q_1$, $\psi_1$, and $Q_2$) for each integration of the transport
equations given above.

\begin{figure}
\epsscale{1.11}
\plotone{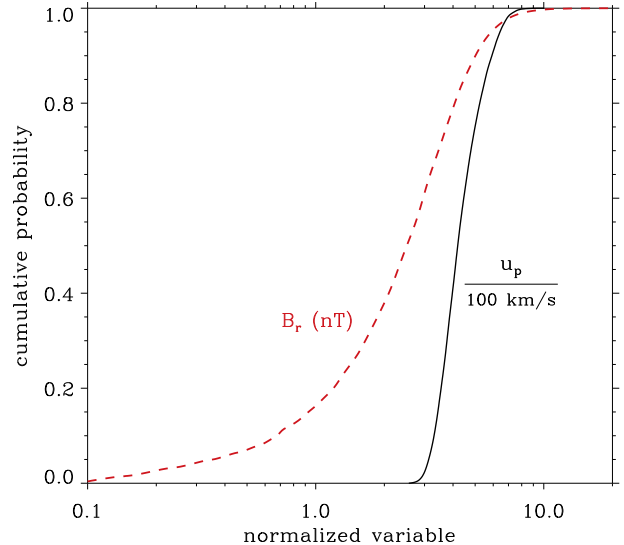}
\caption{Cumulative probability distributions extracted from
OMNI data at 1~AU, where the abscissa corresponds to either
$B_r$ in units of nT (red dashed curve) or $u_p$ in units of
100 km s$^{-1}$ (solid black curve).
\label{fig15}}
\end{figure}

The proton wind speed and radial field strength at 1~AU were sampled
independently from probability distributions constructed from 30 years
(1980 to 2010) of daily averages from the OMNI in~situ dataset
\citep[see, e.g.,][]{KP05}.
Figure \ref{fig15} shows these cumulative distributions.
An individual sample is chosen by computing a pseudorandom number
\citep{PM88} uniformly distributed between 0 and 1, treating it
as the ordinate axis value in Figure \ref{fig15}, and mapping down
to the appropriate abscissa value.
For $u_{1{\rm AU}}$ the minimum, median, and maximum values of the
distribution are 256, 418, and 1003 km s$^{-1}$.
For $B_{r,1{\rm AU}}$ the minimum, median, and maximum values
are 0.1, 2.51, and 22.9 nT.

In the ecliptic-plane OMNI data, the mass flux $\dot{M}$ did not
appear to be strongly correlated with any other parameters.
However, it does vary by about an order of magnitude between values
of $7 \times 10^{-15}$ and $6 \times 10^{-14}$ $M_{\odot}$ yr$^{-1}$.
Thus, we assumed a constant probability distribution in $\log \dot{M}$
and sampled it with a pseudorandom number uniformly distributed
between $-14.15$ and $-13.22$.
In combination with the randomly selected value of $u_{1{\rm AU}}$,
Equation (\ref{eq:rho1au}) provides a realistic distribution of
densities at 1~AU.
Initially, we experimented with randomly choosing the lower boundary
conditions on $T_{p \parallel}$ and $T_{p \perp}$ from a distribution
of likely coronal temperatures, but the results were similar to
what we found with the standard boundary condition described above
($T_{p \parallel} = T_{p \perp} = T_{e}$).

It should be emphasized that our Monte Carlo procedure for selecting
solar wind conditions is only a simplistic first attempt to
``canvass'' the full plasma parameter space.
Our cumulative probability distributions did not explicitly exclude
OMNI data taken during periods when interplanetary coronal mass
ejections (ICMEs) were detected at 1~AU.
Also, we did not consider the effects of stream--stream interactions
in the ecliptic plane, which can produce plasma properties at 1~AU
distinct from those that would otherwise be found in completely
isolated flux tubes \citep[see, e.g.,][]{Bu83,My88,RL11,CvB13}.
Future attempts to compute more accurate heliospheric statistics
will have to take both of these effects into account.

\begin{figure}
\epsscale{1.13}
\plotone{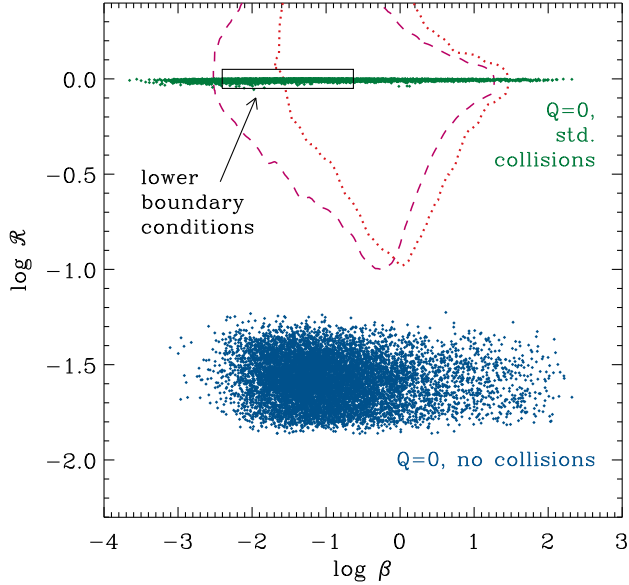}
\caption{Distributions of proton parameters at 1~AU in models with
no heating and randomly varied $u_{1{\rm AU}}$,
$B_{r,1{\rm AU}}$, and $\dot{M}$.
Models with standard Coulomb collisions (green points) and without
collisions (blue points) are compared to the range of initial
values of $\beta$ (black box) and {\em Wind} data boundaries at 1~AU
(red and magenta curves).
\label{fig16}}
\end{figure}

Prior to varying the heating rate parameters, we first show results
of varying $u_{1{\rm AU}}$, $B_{r,1{\rm AU}}$, and $\dot{M}$ while
fixing $Q_{1} = Q_{2} = 0$.
For this set of ``no-heating'' models, the firehose instability
threshold of Equation (\ref{eq:fire}) was not used.
Figure \ref{fig16} shows the conditions at 1~AU for $10^4$ random
models with Coulomb collisions, and $10^4$ models without
collisions.
The same green/blue color scheme from Figure \ref{fig13} is
used in Figure \ref{fig16}.
For both sets of models, the random variations in density and magnetic
field strength gave rise to a distribution of lower boundary
conditions on $\beta$.
Most of the $\beta$ values at $r = 2 \, R_{\odot}$ fell between
0.004 and 0.23, which were the 5\% and 95\% percentile values of
the distribution, respectively.
The heliospheric evolution from 2 $R_{\odot}$ to 1~AU gave rise
to ranges in $\beta$ that are reasonably close to what is observed,
despite the fact that the distributions of ${\cal R}$ values were
clearly unrealistic.

\begin{figure}
\epsscale{1.16}
\plotone{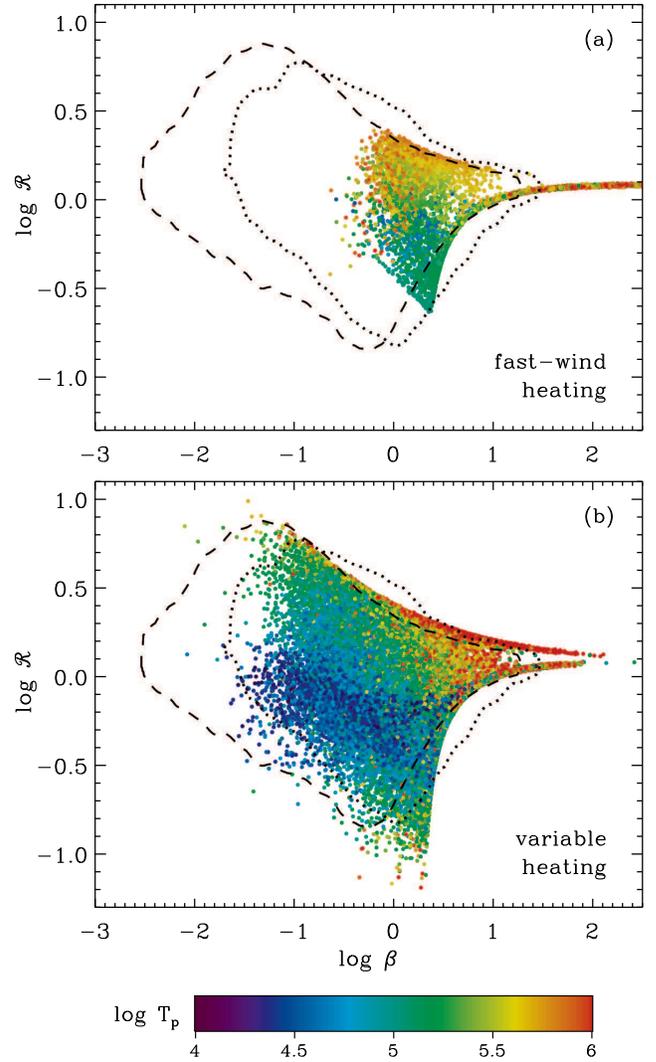}
\caption{Symbols show model results at 1~AU in the beta--anisotropy
plane for random variations of solar wind parameters, with
color corresponding to $T_p$ at 1~AU (see color bar).
In both panels, the {\em Wind} data boundaries at 1~AU are shown
with black dotted and dashed curves.
Models were computed for (a) $Q_p$ fixed at the strongly heated
fast-wind values from Section \ref{sec:radial:result1},
and (b) randomly selected $Q_p$ parameters (see text).
\label{fig17}}
\end{figure}

Next, Figure \ref{fig17}(a) shows the results of $10^4$ trials in
which $u_{1{\rm AU}}$, $B_{r,1{\rm AU}}$, and $\dot{M}$ were varied
randomly and the heating parameters were fixed at values discussed
above for the strongly heated fast wind:
$Q_{1} = 1.596 \times 10^{-6}$ erg s$^{-1}$ cm$^{-3}$,
$\psi_{1} = 4.150$, and
$Q_{2} = 1.660 \times 10^{-6}$ erg s$^{-1}$ cm$^{-3}$.
The symbol color is proportional to the resulting value of $T_p$
at 1~AU, similar to Figure 2 of \citet{Ma11}.
Unlike the more idealized sets of points shown in Figure \ref{fig14},
in this case the random variation of background wind conditions
produces a significant ``filled-in'' region of the
($\beta$,${\cal R}$) diagram at 1~AU.
However, only the right-hand side of the diagram is populated by
the models (i.e., mainly $\beta \gtrsim 0.3$), since the fast-wind
values of $Q_1$ and $Q_2$ represent rather strong rates of heating.
It appears that only weaker heating rates will allow the left-hand
side of the diagram to become filled in.
The occupied part of parameter space in Figure \ref{fig17}(a) also
resembles plots of measurements restricted to the fast wind, such
as Figure 2 of \citet{Hg06}.

Figure \ref{fig17}(b) shows the result of a full Monte Carlo
simulation that involved varying all six parameters of the model.
In addition to randomly choosing $u_{1{\rm AU}}$, $B_{r,1{\rm AU}}$,
and $\dot{M}$, the proton heating rates were sampled from random
distributions as well.
$Q_1$ and $Q_2$ were each sampled uniformly on a logarithmic scale
between $10^{-8}$ and $10^{-5}$ erg s$^{-1}$ cm$^{-3}$, and the
value of $Q_{\rm cyc}$ at 1~AU was sampled uniformly between
$10^{-17}$ and $10^{-14}$ erg s$^{-1}$ cm$^{-3}$.
This produced a distribution in $\psi_1$ exponents that ranged
between 2.57 and 5.14, and had a mean value of 3.85.
Vertical bars in Figure \ref{fig10} illustrate the ranges of
variation of the heating rates at $r = 2 \, R_{\odot}$ and 1~AU.
Even though the heating rates are varied over three
orders of magnitude, these ranges never stray too far from
existing expectations from other models and observations.

The full Monte Carlo model was run for 95,000 random trials.
However, we found that roughly a third of these trials ended up
with somewhat unrealistic conditions at 1~AU. 
Thus, Figure \ref{fig17}(b) shows only approximately 60,000 of them.
Unrealistic models were excluded based on two criteria:
\begin{enumerate}
\item
The largest collisional ages at 1~AU were found to correspond
to models with unusually weak heating.
Their heliospheric trajectories resembled the green dotted
curves of Figure \ref{fig13}.
Thus, we eliminated roughly 20,000 models that exhibited
$A_{p} > 2$ at 1~AU.
\citet{Ma12} also made a similar collisional age cut to the
{\em Wind} data in order to better isolate cases of collisionless
wave-particle interaction.
For the eliminated models, the mean value of $T_p$ at 1~AU was only
7,400 K, which is significantly smaller than the minimum proton
temperature of $\sim$12,000 K seen in the OMNI data.
\item
Even though the heating rates were not assumed to scale with the
radial magnetic field strength, we do expect the solar wind to
exhibit some kind of magnetic correlation
\citep[see, e.g.,][]{Pi09,Cr09,Wa10,WC14}.
Thus, it was not surprising to see that when the largest values
of $Q_2$ were paired with the smallest values of $B_r$, the
proton heating at 1~AU was unusually strong.
This effect can be parameterized by computing the rate (i.e.,
inverse timescale) $\nu_{\rm B} = Q_{\rm iso}/U_{\rm B}$ at 1~AU,
where $U_{\rm B} = B_{r}^{2}/8\pi$ is the magnetic energy density.
Let us also define a normalizing value of $\nu_{\rm B}$ that
corresponds to the strongly heated fast-wind model of
Section \ref{sec:radial:result1}; this value is given by
$\nu_{{\rm B}\ast} = 1.48 \times 10^{-6}$ s$^{-1}$.
The full grid of 90,000 models contained cases with
$\nu_{\rm B} / \nu_{{\rm B}\ast}$ ranging over six orders of
magnitude between $10^{-3}$ and $10^{+3}$.
We eliminated roughly 15,000 models that exhibited
$\nu_{\rm B} / \nu_{{\rm B}\ast} > 5$.
For that subsample, the mean values of $T_p$ and $\beta$ at 1~AU
were 1.6 MK and 950, which are well above the maximum values
from OMNI of $\sim$0.4 MK and $\sim$30, respectively.
\end{enumerate}

Applying the above criteria gives rise to the $\sim$60,000 models
shown in Figure \ref{fig17}(b).
The area of occupied space in the beta--anisotropy plane
agrees well with observations at 1~AU.
Also, the correlation between $T_p$ (shown with symbol color) and
location in the diagram is similar to that in the {\em Wind} data
of \citet{Ma11}.
The {\em left edges} of the occupied region correspond to the
lowest values of $\nu_{\rm B}$ at 1~AU.
For those models, there is not enough heating
``per unit field strength'' to drive the wind to high $\beta$.
Whether a weakly heated model winds up on the upper or lower
left edge depends on the extent to which the heating is mostly
isotropic or mostly cyclotron.
At the left edge (i.e., for $\beta < 0.1$), we found that the
anisotropy ratio ${\cal R}_{\rm left}$ is strongly correlated with
the heating ratio $Q_{\rm cyc}/Q_{\rm iso}$ at 1~AU, and
\begin{equation}
  {\cal R}_{\rm left} \, \approx \, 0.15 \sqrt{ \left(
  \frac{Q_{\rm cyc}}{Q_{\rm iso}} \right)_{1 \, {\rm AU}} }
  \,\, .
\end{equation}
Thus, if this model captures the real physics of the solar wind,
it requires there to be a broad range of partition fractions
between the two types of heating.
This range must also extend up to
$Q_{\rm cyc}/Q_{\rm iso} \gtrsim 100$ in order to produce the
upper-left edge of the diagram.

\begin{figure}
\epsscale{1.11}
\plotone{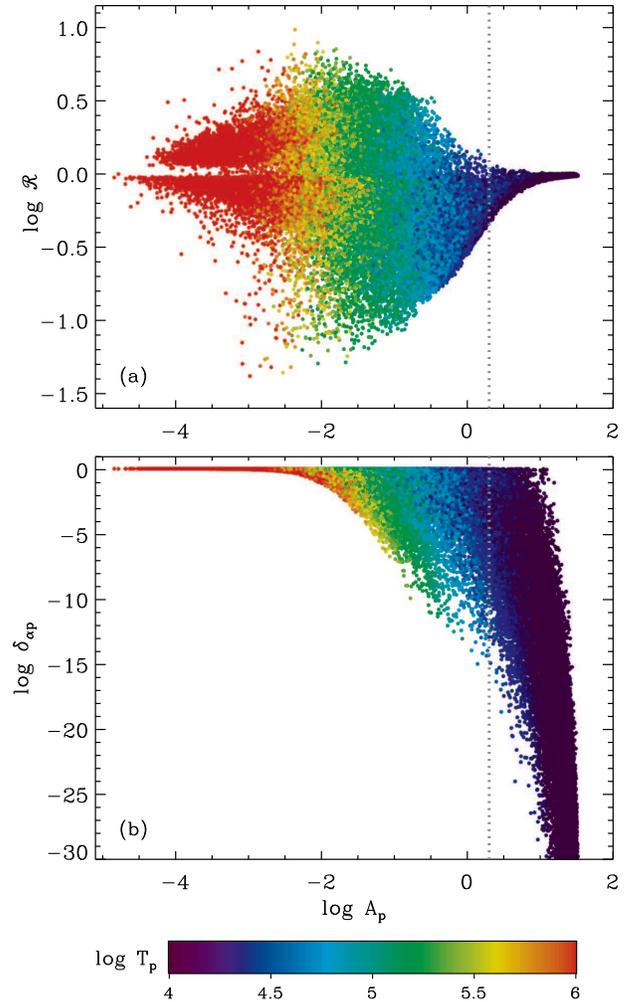}
\caption{Dependence of: (a) anisotropy ratio ${\cal R}$,
(b) alpha--proton drift parameter $\delta_{\alpha p}$ on the
collisional age $A_p$.
Symbol colors are proportional to $T_p$ at 1 AU (see color bar).
The gray dotted line illustrates the cutoff at $A_{p}=2$ applied
to models shown Figure \ref{fig17}.
Note that many of the ``coldest'' models to the right of the
cutoff have $\log T_p$ that extends down to $\sim$3.5, but
the color scale saturates at $\log T_{p} = 4$.
\label{fig18}}
\end{figure}

Figure \ref{fig18} illustrates how the Monte Carlo models
behave in terms of their Coulomb collisional coupling and
relative alpha--proton drift.
Here, we applied only the second of the two above criteria (i.e.,
eliminating models with $\nu_{\rm B} / \nu_{{\rm B}\ast} > 5$) and
we show the full range of collisional ages.
The inverse temperature dependence in the collision rate is evident
in that $A_{p} \propto T_{p}^{-3/2}$ \citep[see also][]{HT14}.
It is clear that the models experiencing the most frequent collisions
are driven toward the lowest values of $\delta_{\alpha p}$ at 1~AU,
and toward ${\cal R} \approx 1$.
However, the {\em least} collisional models (i.e.,
$A_{p} \lesssim 10^{-4}$) also seem to be driven toward isotropy.
These models have large values of $\beta$ at 1~AU and thus are
constrained in ${\cal R}$ because of the cyclotron and firehose
instability boundaries.

\section{Conclusions and Discussion}
\label{sec:conc}

The goal of this paper was to begin exploring the detailed kinetic
consequences of an assumed population of ion cyclotron resonant
waves on the large-scale evolution of the solar wind.
We found that the full range of kinetic properties of protons
seen at 1~AU is generally reproducible with a stochastic ensemble
of both background wind parameters and cyclotron heating rates.
Other recent simulations \citep{Sg13,Sv14} have come to similar
conclusions, but their random spreads of parameter choices were
not so clearly connected to the actual ``histories'' of the
modeled solar wind parcels.
We also found that a broad range of {\em relative contributions}
of ion cyclotron heating to the total heating rate (i.e., a broad
range of $Q_{\rm cyc}/Q_{\rm iso}$ ratios) may be important to
reproducing the low-$\beta$ part of observed parameter space.

In the process of building models of proton VDF transport from
the corona to 1~AU, some additional key physical properties of
ion cyclotron resonance were revealed.  For example,
\begin{enumerate}
\item
In Section \ref{sec:disp:bimax}, a two-term asymptotic expansion 
of the plasma dispersion function was used to derive a new
analytic ``warm plasma'' dispersion relation that simultaneously
deals with both ion cyclotron resonance and the firehose-unstable
regime.
This dispersion relation (see Equation (\ref{eq:wad}) and
Figure \ref{fig01}(b)) also reproduces much of the $\beta$
and anisotropy dependence found in numerical Alfv\'{e}n-wave
solutions of the full Vlasov-Maxwell dispersion relation.
\item
In Section \ref{sec:disp:anshell}, we found analytic expressions
for a modified plasma dispersion function consistent with
a simple form of a non-Maxwellian ``resonant shell'' proton VDF.
Although the resulting dispersion relation was found to be
equivalent to the two-term expansion discussed above,
this derivation showed that some results coming from bi-Maxwellian
theory may still be applicable to other types of kinetic
distributions.
\item
Section \ref{sec:heat:diff} described a generalization of
quasilinear theory that replaces the Dirac delta function
associated with wave-particle resonance with a more realistic
Lorentzian function that depends on the damping rate $\gamma$.
This expression leads to nonzero VDF diffusion coefficients in
regions of velocity space that are classically nonresonant (e.g.,
$v_{\parallel} > 0$ for the resonance between protons and
forward-propagating ion cyclotron waves).
\item
The numerical VDF diffusion models of \citet{Cr01} were extended
to protons in the presence of both forward- and backward-propagating
Alfv\'{e}n waves.
The early-time evolution of proton temperature moments was found
to be nearly identical to equivalent bi-Maxwellian calculations.
Also, it appeared that the late-time evolution into non-Maxwellian
shell-type VDFs exhibits cross-shell perpendicular diffusion
\citep[see, e.g.,][]{Is01} even for purely forward-propagating waves.
\item
In Section \ref{sec:heat:bimax}, we solved for the locations of
marginal stability curves (defined by a vanishing time derivative
of the proton anisotropy ratio) in the beta--anisotropy plane
for several different dispersion relations.
Figures \ref{fig04}(b) and \ref{fig09} displayed a surprising
diversity of curve shapes.
This indicates that any conclusions made about the ion cyclotron
instability boundaries may be extremely sensitive to assumptions
made about the wave dispersion and particle VDFs.
\item
Section \ref{sec:alpha} outlined a damped cascade model that
describes how drifting alpha particles may intercept ion
cyclotron waves prior to them becoming resonant with protons.
Figure \ref{fig12} shows how this model appears able to reproduce
an empirical dividing line in ($\beta$, $\delta_{\alpha p}$)
space seen in {\em Wind} data \citep{Kp13}.
\end{enumerate}

Despite the new insights gained from the exploratory modeling
described above, it did not completely reproduce all observational
constraints.
At any rate, it points the way to improved methods of simulating
the radial evolution of a realistic ensemble of solar wind states.
Other proposed explanations for preferential proton heating in
the heliosphere---such as stochastic KAW heating
\citep{Cs10,Ch11,Ch13,BC13}---can be straightforwardly plugged into
the same kind of transport model as the one described in
Section \ref{sec:radial}.

In addition to the need for testing other physical processes, it is
also clear that the existing models of ion cyclotron resonance need
to be improved in several important ways.
This paper relied heavily on the bi-Maxwellian assumption.
One of our goals was in fact to explore in detail how
bi-Maxwellians can be useful tools for understanding the kinetic
physics of the solar wind.
However, it is well-known that measured VDFs often exhibit
marked departures from this simple parameterization.
Other kinetic simulation approaches
\citep{IV11,GS13,Aj13,Mn14,Sv14} are helping to make clear how
specific departures from the bi-Maxwellian model affect the
conclusions made about the large-scale energy balance of the
heliosphere.

The models described in this paper were also limited to purely
parallel-propagating Alfv\'{e}n waves.
However, the solar wind fluctuation spectrum is known
to exhibit a broad range of propagation angles.
There have been several studies of the radial evolution of
this angular distribution \citep{Bv82,Hb95,N04,He13}.
Although there is still no agreement on the overall sense
of the evolution---i.e., is it becoming more parallel or
more perpendicular with increasing distance?---it is evident
that there must be wave power at all values of $\theta$.
Obliquely propagating Alfv\'{e}n/ion-cyclotron waves have been
studied extensively in a solar wind context
\citep[see, e.g.,][]{LH01,HC05,Ch10,LL10,He12}, and some
properties of the particle heating are different from the
parallel-propagating case.

Understanding the proton thermodynamics might also depend on
including couplings between Alfv\'{e}n waves and other
compressive MHD modes
\citep[see][and references therein]{CvB12,Mn14}.
One possibly important example, not studied in this paper, is
the mirror-mode wave.
It has been claimed that the upper-right boundary in the beta--anisotropy
diagram is better described by the mirror instability than by the
cyclotron instability \citep[e.g.,][]{Hg06,Ba09,Ma12}.
Mirror-mode waves have been detected sporadically in the solar wind
\citep{Ru09}, and it is possible that the well-known pressure-balanced
structures (PBSs) may be caused by them \citep{Ya13}.
It is interesting to note that \citet{Lb81} modeled solar wind ion
VDFs using a skewed non-Maxwellian function meant to be consistent
with a nonzero heat flux, and found that the mirror instability
completely disappears, leaving only the ion cyclotron instability
for ${\cal R} > 1$.

Lastly, the properties of the alpha particles should be modeled
simultaneously with the protons, and with a comparable degree of
physical realism \citep[see, e.g.,][]{Bu13,Ch13,My14}.
At the very least, the sensitivity of the results in Sections
\ref{sec:alpha}--\ref{sec:radial} to our assumption that
$T_{\alpha}/T_{p} = 4$ needs to be explored.
It would also be beneficial for models to make predictions of the
detailed plasma properties of protons and alpha particles at a
range of distances that extends down to $\sim$9 $R_{\odot}$,
the latter being the inner boundary of future exploration by
{\em Solar Probe Plus} \citep{Fx13}.

\acknowledgments

The author gratefully acknowledges Adriaan van Ballegooijen,
Lauren Woolsey, Phil Isenberg, Peter Gary, Justin Kasper, and
Ben Maruca for many valuable discussions.
This work was supported by NASA grant {NNX\-10\-AC11G}
and NSF SHINE program grant AGS-1259519.
The OMNI solar wind data were obtained from the NASA/GSFC
Space Physics Data Facility's OMNIWeb service, and we thank the
principal investigators of the {\em IMP 8}, {\em Wind,} and
{\em ACE} instruments who provided their data to OMNI.
This research made extensive use of NASA's Astrophysics Data
System (ADS).

\end{document}